\documentclass[11pt]{article}

\newcommand{\lncsonly}[1]{}
\newcommand{\articleonly}[1]{#1}
\newcommand{\acmonly}[1]{}
\newcommand{\ijcaionly}[1]{}
\newcommand{\ieeeonly}[1]{}

\newcommand{\lfcsonly}[1]{}
\newcommand{\fullonly}[1]{#1}
\usepackage{comment}
\newif\iffullversion
\fullversiontrue    %
\iffullversion
  \newenvironment{fullversion}{}{}
\else
  \excludecomment{fullversion}
\fi

\lncsonly{}

\articleonly{
\usepackage{lipsum}
\let\OLDthebibliography\thebibliography
\renewcommand\thebibliography[1]{
  \OLDthebibliography{#1}
  \setlength{\parskip}{1.5pt}
  \setlength{\itemsep}{0pt plus 0.3ex}
}
\usepackage{fullpage}
\usepackage{graphicx}
\usepackage{amssymb}  %
\fullonly{\bibliographystyle{alpha_firstinit}}
\lfcsonly{}
}

\acmonly{}

\ijcaionly{}

\ieeeonly{}

\usepackage{url}
\usepackage{alltt} %
\usepackage[T1]{fontenc} %
\usepackage{xspace} 
\usepackage{lmodern} 

\fullonly{\usepackage[subtle,mathspacing=normal]{savetrees}} %
\lfcsonly{} %

\newcommand{\Hex}[1]{\hspace{#1ex}}
\newcommand{\Vex}[1]{\vspace{#1ex}}

\usepackage{tikz}
\usepackage{pgfplots}
\pgfplotsset{scaled x ticks=false}
\pgfplotsset{compat=1.14} %

\usepackage{enumitem}
\setlist{itemsep=0ex, topsep=1ex, parsep=1ex, leftmargin=4ex}

\newcommand{\mysec}[1]{\Vex{0}\section{#1}} %
\newcommand{\mysubsec}[1]{\Vex{-0}\subsection{#1}\Vex{-1}}
\newcommand{\mypar}[1]{\Vex{2}\noindent {\bf #1.}~}

\newenvironment{code}{\Vex{0}\begin{alltt}\small}{\end{alltt}\Vex{0}} %
\newcommand\co[1]{{\lfcsonly{}\small\tt #1}} %
\newcommand\mco[1]{\mbox{\footnotesize\tt #1}}
\def\mathify#1{\ifmmode{#1}\else\mbox{$#1$}\fi}
\newcommand\m[1]{\mathify{#1}}  %

\newcommand\IF{\m{\leftarrow}\xspace}
\newcommand\AND{\m{\land}\xspace}
\newcommand\OR{\mathify{\lor}\xspace}
\newcommand\NOT{\m{\neg}\xspace}
\newcommand\IMPLIES{\m{\to}\xspace}
\newcommand\SOME{\mathify{\exists}\xspace}
\newcommand\EACH{\mathify{\forall}\xspace}

\newcommand\SUBSET{\mathify{\subseteq}\xspace}
\newcommand\EQ{\mathify{=}\xspace}
\newcommand\NEQ{\mathify{\neq}\xspace}
\newcommand\GEQ{\mathify{\geq}\xspace}
\newcommand\LEQ{\mathify{\leq}\xspace}
\newcommand\GT{\mathify{>}\xspace}
\newcommand\LT{\mathify{<}\xspace}
\newcommand\T{\mathify{\it True}\xspace}
\newcommand\F{\mathify{\it False}\xspace}
\newcommand\UD{\mathify{\it Undef}\xspace}

\newcommand{\notes}[1]{} %

\newcommand\fund{
    This work was supported in part by NSF under grants 
    CCF-1954837, %
    CCF-1414078, %
    and IIS-1447549 %
    and by ONR under grants
    N00014-20-1-2751 %
    and N00014-21-1-2719. %
    }

\begin{document}
\title{\lncsonly{}%
\articleonly{\lfcsonly{} Recursive Rules with Aggregation:
A Simple Unified Semantics}%
\acmonly{}%
\thanks{\fund}
} %
\lncsonly{}
\articleonly{
\author{Yanhong A. Liu \Hex{10} Scott D. Stoller\\
\fullonly{\normalsize
Computer Science Department, Stony Brook University, Stony Brook, New York\\}
\co{liu@cs.stonybrook.edu \Hex{\fullonly{6}\lfcsonly{}} stoller@cs.stonybrook.edu\Hex{-3}}
}}
\acmonly{}
\ijcaionly{}
\ieeeonly{}
\fullonly{\date{}}%
\lfcsonly{}

\newcommand{\abstracttext}{%

Complex reasoning problems are most clearly and easily specified using
logical rules, but require recursive rules with aggregation such as
count and sum for practical applications.  Unfortunately, the meaning of
such rules has been a significant challenge, leading to many disagreeing
semantics.

This paper describes a unified semantics for recursive rules with
aggregation, extending the unified founded semantics and constraint
semantics for recursive rules with negation.  The key idea is to support
simple expression of the different assumptions underlying different
semantics, and orthogonally interpret aggregation operations
using their simple usual meaning.
We present a formal definition of the semantics, prove important
properties of the semantics, and compare with prior semantics.
In particular, we present an efficient inference over aggregation
that gives precise answers to all examples we have studied from the literature.  
We also \fullonly{apply}\lfcsonly{} our semantics to a wide range of challenging examples,
\lfcsonly{}\fullonly{and show that our semantics 
is simple and matches the desired results in all cases.
Finally, we describe experiments on the most challenging examples,
exhibiting unexpectedly superior performance over well-known systems 
when they can compute correct answers. 
}\lfcsonly{}}

\acmonly{}

\maketitle

\lncsonly{}
\articleonly{\begin{abstract}
\abstracttext
\end{abstract}}
\ijcaionly{}
\ieeeonly{}

\mysec{Introduction}
\label{sec-intro}

Many computation problems, including complex reasoning problems in
particular, such as program analysis, networking, and decision support, are most clearly and easily specified using logical rules~\cite{liu18LPappl-wbook}.
However, such reasoning problems in practical applications, especially for
large applications and when faced with uncertain situations, require the
use of recursive rules with aggregation such as count and sum.
Unfortunately, the meaning of such rules has been challenging and
remains a subject with significant complication and disagreement.

As a simple example, consider a single rule for Tom to attend the logic
seminar: ``Tom will attend the logic seminar if the number of people who will
attend it is at least 20.''  What does the rule mean?  If 20 or more other
people will attend, then surely Tom will attend.  If only 10 others will
attend, then Tom will not attend.  What if only 19 other people
will attend?  Will Tom attend, or not?
Although simple, this example already shows that, when aggregation is used
in recursive rules---here count is used in a rule that defines "will attend" using "will attend"---the semantics can be tricky.

Some might say that this statement about Tom %
is ambiguous or ill-specified.  However, it is a statement allowed by
logic rule languages with predicates, sets, and counts. For example,
let predicate \co{will\_attend(p)} denote that \co{p} will attend 
the logic seminar; then the statement can be written as 
\co{will\_attend(Tom) if count \{p:~will\_attend(p)\} \GEQ 20}.  
So the statement must be given a meaning.
Indeed, ``ambiguous'' is a possible meaning, indicating there are two or more answers, and ``ill-specified'' is another possible meaning, indicating there is no answer. %
So which one should it be?  Are there other possible meanings?

In deductive databases, to avoid challenging cases of aggregation as well as negation, processing of recursive rules with aggregation is largely limited to monotonic programs, 
i.e., adding a new fact used in a hypothesis 
cannot make a conclusion change from true to false.  However, note that the rule about Tom attending the logic seminar is actually monotonic: 
adding \co{attend(p)} for a new \co{p}
can not make the conclusion change from true to false.
So, even restricted deductive databases must give a meaning to this rule.  What should it be?

In fact, the many different semantics of recursive rules with aggregation are
more complex and trickier than even semantics of recursive rules with negation.  The
latter was already challenging for over 120 years, going back at least to
Russell's paradox, for which self-reference with negation is believed to
form vicious circles~\cite{irvine20russell}.  Many different semantics,
which disagree with each other, have been studied for recursive rules with
negation, as summarized in Section~\ref{sec-related}.
Two of them, well-founded semantics (WFS)~\cite{van+91well,van93alt} and
stable model semantics (SMS)~\cite{GelLif88}, became dominant since about
30 years ago.

Semantics of recursive rules with aggregation has been studied continuously
since about 30 years ago, and more intensively in more recent years, as
discussed in Section~\ref{sec-related}, especially as they are needed in
graph analysis and machine learning applications.  However, the many different
semantics proposed, e.g.,~\cite{van92agg,gelfond2019vicious},
are even more intricate than WFS and SMS for recursive rules with negation,
and include even different extensions for WFS, e.g.,~\cite{kemp1991semantics,van92agg,sudarshan1993extending}, and for SMS, e.g.,~\cite{kemp1991semantics,marek2004set,gelfond2019vicious}.
Some authors also changed their own minds about the desired semantics,
e.g.,~\cite{gelfond2002representing,gelfond2019vicious}.  
Such intricate and disagreeing semantics would be too challenging to use correctly.

This paper describes a simple unified semantics for recursive rules with
aggregation as well as negation and quantification.  The semantics is
built on and extends the founded semantics and constraint semantics of
logical rules with negation and quantification developed recently by Liu
and
Stoller~\cite{LiuSto18Founded-LFCS,LiuSto20Founded-JLC}.
The key idea is to capture, and to express in a simple way, the different assumptions
underlying different semantics, and orthogonally interpret aggregation
operations %
using their simple usual meaning.
We present formal definition of the semantics
and prove important properties of the semantics. \fullonly{

}In particular, we present an efficient derivability relation for comparisons containing aggregations; it can be computed in linear time and gives precise answers on all examples we have studied %
from the literature. 
\lfcsonly{}We also compare\lfcsonly{} with main prior semantics for rules with aggregation, and show\lfcsonly{} how our semantics is direct and follows precisely from usual meanings of aggregations. \fullonly{ 

}We further \fullonly{apply}\lfcsonly{} our semantics to 
a wide range of challenging examples, and show\lfcsonly{} that
our semantics is simple and matches the desired semantics in all cases.
\fullonly{In particular, our examples include the longest and most sophisticated ones 
from dozens of previously studied examples~\cite{gelfond2019vicious},
where answer sets were obtained by all possible guesses %
followed by computing sophisticated reduct for each;
we show that all the answer sets can be obtained by
a simple default assumption and simple least fixed-point computations, as
is usual for inductive definitions and commonsense reasoning.

Finally, we describe experiments on three most challenging examples.
For one of them, our systematically incrementalized implementation in Python exhibits unexpectedly superior performance over four other most well-known systems. For the other two, we discovered many incorrect results computed by two systems, while the other two systems cannot compute the desired semantics at all.
The former two systems have since found and fixed bugs that caused the
incorrect results in one of the examples, but developers of these systems also found that
fundamentally different inference would be needed to compute correctly at
all for the other example.}\lfcsonly{}
\lfcsonly{}

\fullonly{
The rest of the paper is organized as follows.  Section~\ref{sec-prob}
presents the problems and an overview of the solutions.
Sections~\ref{sec-lang} describes the language of recursive rules with
unrestricted negation, quantification, and aggregation.
Sections~\ref{sec-sem}\lfcsonly{}\fullonly{ and \ref{sec-props} present}
formal definition of the semantics and prove
properties of the semantics.\fullonly{
Section \ref{sec-cmp} discusses and compares with prior semantics.}
Section~\ref{sec-examp}
illustrates our semantics on a wide range of examples.\fullonly{
Section~\ref{sec-expe} describes experimental results.}
Section~\ref{sec-related} discusses related work and concludes.\fullonly{
Appendix~\ref{sec-proofs} contains additional proofs.}

This paper is an extended and revised version of Liu and
Stoller~\cite{LiuSto22RuleAgg-LFCS}.  The main changes are 
(1) new Section~\ref{sec-props} on properties of the semantics, extending a
summary paragraph to include all theorems with proofs, including additional
proofs in Appendix~\ref{sec-proofs};
(2) new Section~\ref{sec-cmp} that discusses and compares with prior
semantics in more detail;
(3) extended Section~\ref{sec-examp} on examples, with additional examples 1-4 
and 7; and
(4) new Section~\ref{sec-expe} on experiments, extending a summary description 
about two examples to full descriptions about three examples.
}

\mysec{Problem and solution overview}
\label{sec-prob}

The semantics of recursion with negation and aggregation is challenging for several reasons.
First, recursion involves self-referencing and cyclic reasoning, for which it is already non-trivial to properly start and finish.
Then, negation in recursion incurs self-denying and conflict in cyclic reasoning, which can lead to contradiction, as in Russell's paradox.
Finally, aggregation generalizes negation to give rise to even greater challenges in recursion, %
because a negation essentially corresponds to only the simple case of a count being zero.

The first reason alone already called for a least fixed point semantics, which is beyond first-order logic. %
The second reason led to major different semantics that are sophisticated and disagreeing when trying to solve conflicts differently.  The third reason exacerbated the sophistication and variety to tackle the even greater challenges.

\mypar{A smallest example}
Consider the following %
recursive rule with aggregation.
It says that \co{p} is true for value \co{a} if the number of \co{x}'s for
which \co{p} is true equals \co{1}:
\begin{code}
  p(a) \IF count \{x: p(x)\} \EQ 1  
\end{code}
This rule is recursive because inferring a conclusion about \co{p} requires
using \co{p} in a hypothesis.  It uses an aggregation of \co{count} over a
set.
While each of recursion and aggregation by itself has a simple meaning,
allowing recursion with aggregation is tricky, because recursion is used to
define a predicate, which is equivalent to a set, but aggregation using a
set requires the set to be already defined.

We use this example in addition to our Tom example in Section~\ref{sec-intro}, for two reasons.  
First, this and similar small examples are used for comparisons in previous papers, e.g.,~\cite{faber2011semantics,gelfond2014vicious,gelfond2018vicious}.
Second, this example differs from the Tom example in that the comparison with count in this example is non-monotonic, i.e., adding more \co{x}'s for which \co{p(x)} is true can change the value of the comparison, and thus the conclusion, from true to false;
using only one example is insufficient to show the main different cases.

\begin{itemize}
\item {\bf Two models: Kemp-Stuckey 1991, Gelfond 2002.}~
According to Kemp and Stuckey~\cite{kemp1991semantics} and
Gelfond~\cite{gelfond2002representing}, the above rule has two models: one
empty model, i.e., a model in which nothing is true and thus \co{p(a)}
is false, and one containing only \co{p(a)}\fullonly{, i.e., \co{p(a)} is true and
everything else is false}\lfcsonly{}.

\item {\bf One model: Faber et al.\,2011, Gelfond-Zhang 2014-2019.}
According to Faber, Pfeifer, and Leone~\cite{faber2011semantics} and
Gelfond and Zhang~\cite[Examples 2 and 7]{gelfond2014vicious},
\cite[Examples 4 and 6]{gelfond2018vicious}, and~\cite[Example
9]{gelfond2019vicious}, the rule above has only one model: the empty model.
\end{itemize}

As one of the several main efforts investigating aggregation, Gelfond and
Zhang~\cite{gelfond2002representing,gelfond2014vicious,zhang2016characterization,gelfond2017vicious,gelfond2018vicious,gelfond2019vicious}
have studied the challenges and solutions extensively, presenting
dozens of definitions and propositions and discussing dozens of
examples~\cite{gelfond2019vicious}.  Their examples where count is used in
inequalities, greater than, etc., with additional variables, with more
hypotheses in a rule, or with more rules and facts, 
are even more complicated. %
\fullonly{
We discuss their most extensive examples in Section~\ref{sec-examp}.}

\mypar{Extending founded semantics and constraint semantics for
  aggregation}
Aggregation, such as count, is a simple concept that even kids understand.
So it is stunning to see so many sophisticated treatments %
for figuring out its meaning when it is used in rules, 
and to see the many disagreeing semantics resulting from those.

We develop a simple and unified semantics for rules with aggregation as
well as negation and quantification by building on founded semantics and
constraint semantics~\cite{LiuSto18Founded-LFCS,LiuSto20Founded-JLC} for
rules with negation and quantification.  The key insight is that
disagreeing complex semantics for rules with aggregation are because of
different underlying assumptions, and these assumptions can be captured
using the same simple binary declarations about predicates as in founded
semantics and constraint semantics but generalized to include the meaning
of aggregation.

\begin{description}
  \setlength{\parskip}{1ex}
  
\item {\bf Certain.}~
First, if there is no potential non-monotonicity, 
including no aggregation in recursion, then the predicate
in the conclusion can be declared ``certain''.  

Being certain means that assertions of the predicate are given true or
inferred true by 
simply following rules whose hypotheses are given or inferred true,
and the remaining assertions of the predicate are false.
This is both the founded semantics and constraint semantics.

{For the Tom example}, there is no potential
non-monotonicity; with this declaration, when given that only 19 others
will attend, the hypothesis of the rule is not true, so the conclusion
cannot be inferred.  Thus Tom will not attend.

\item {\bf Uncertain.}~
Regardless of monotonicity, a predicate can be declared ``uncertain''.  
It means that assertions of the predicate can be given or inferred true or
false using what is given, %
and any remaining assertions of the predicate
are undefined.  This is the founded semantics.

If there are undefined assertions from founded semantics, all combinations of
true and false values are checked against the rules and declarations as constraints,
yielding a set of possible satisfying combinations.  This is the constraint
semantics.

\item {\bf Complete, or not complete.}~
An uncertain predicate can be further declared ``complete'' or not.  

Being complete means that all rules that can conclude assertions of the predicate
are given.  Thus a new rule, called completion rule, can be created to
infer negative assertions of the predicate when none of the given rules
apply.

Being not complete means that negative assertions cannot be inferred using
completion rules, and thus all assertions of the predicate that were not
inferred to be true are undefined.

For the Tom example, the completion rule implies: 
Tom will not attend the logic seminar if the number of
people who will attend it is less than 20.

When given that only 19 others will attend, due to the
uncertainty of whether Tom will attend, neither the given rule nor the
completion rule will fire.
So whether one uses the declaration of complete or not, there is no way to
infer that Tom will attend, or Tom will not attend.  So, founded semantics
says it is undefined.

Then constraint semantics tries both for it to be true, and for it to be
false; both satisfy the rule, so there are two models: one where Tom will
attend, and one where Tom will not attend.

\item {\bf Closed, or not closed.}~
Finally, an uncertain complete predicate can be further declared 
``closed'' or not.

Being closed means that an assertion of the predicate is made false if
inferring it to be true requires itself to be true.

Being not closed means that such assertions are left undefined.

For the Tom example, with this declaration,
if there are only 19 others attending, then Tom will not attend in both
founded semantics and constraint semantics.  This is because inferring that
Tom will attend requires Tom himself to attend to make the count to be 20,
so it should be made false, meaning that Tom will not attend.

Note that this is the same result as using "certain".
Because the rule for deciding whether Tom will attend has no potential
non-monotonicity, 
using ``certain'' is much simpler and has the same meaning as using
``closed'', as \fullonly{shown}\lfcsonly{} in general in Section~\ref{sec-props}.
\end{description}

For the smallest example about \co{p} near the beginning of this section, the equality
comparison is not monotonic.  Thus \co{p} must be declared uncertain.  This example also shows different semantics when using the declarations of not complete and complete, unlike the Tom example.
\begin{itemize}

\item {\bf Not complete.}~ 
Suppose \co{p} is declared not complete.  Founded semantics does not
  infer \co{p(a)} to be true using the given rule because \co{count
    \{x:~p(x)\} = 1} cannot be determined to be true, and nothing infers \co{p(a)} to be
  false.  Thus \co{p(a)} is undefined.  So is \co{p(b)} for any constant
  \co{b} other than \co{a} because nothing infers \co{p(b)} to be true or false.
  Constraint semantics gives a set of models, each for a different
  combination of true and false values of \co{p(c)} for different constants \co{c} such that the combination satisfies the given rule.
  This corresponds to what is often called open-world assumption and used
  in commonsense reasoning.

\item {\bf Complete.}~
Suppose \co{p} is declared complete but not closed. A completion rule
  is first added.  The precise completion rule is:
\begin{code}
  \NOT p(x) \IF x \NEQ a \OR count \{x: p(x)\} \NEQ 1
\end{code}
  Founded semantics does not infer \co{p(a)} to be true or false
  using the given rule or completion rule, because 
  \co{count \{x:~p(x)\} \NEQ 1} also
  cannot be determined to be true.  Thus \co{p(a)} is
  undefined.  Founded semantics infers \co{p(b)} for any constant \co{b}
  other than \co{a} to be false using the completion rule.
  Constraint semantics gives two models: one with \co{p(a)} being true, and
  \co{p(b)} being false for any constant \co{b} other than \co{a}; and one
  with \co{p(c)} being false for every constant \co{c}.  
  This is the same as the two-model
  semantics per Kemp-Stuckey~1991 and~Gelfond 2002.

\item {\bf Closed.}~  Supposed \co{p} is declared %
  complete and closed.  Both founded semantics and constraint semantics
  give only the second model above, i.e., \co{p(c)} is false for every
  constant \co{c}.  They have \co{p(a)} being false because inferring
  \co{p(a)} to be true requires \co{p(a)} itself to be true.  
  This is the same as the one-model semantics per Faber et al.~2011
  and Gelfond-Zhang~2014-2019.
\end{itemize}

We see that simple binary declarations of the underlying assumptions, with
simple inference following rules and taking rules as constraints, give the
different desired semantics.

\mypar{Relationship with prior semantics}
Table~\ref{tab-FSCS} summarizes relationships between our unifying semantics 
and major prior semantics.
With different predicate declarations capturing different underlying assumptions,
founded semantics and constraint semantics for rules with aggregation extend different prior semantics for rules with negation uniformly, as shown in Table~\ref{tab-FSCS} left and middle columns.
These extend the matching relationships proved for rules with negation in~\cite{LiuSto18Founded-LFCS,LiuSto20Founded-JLC}.
All these relationships are when all predicates in a program have the same declarations, but our founded semantics and constraint semantics also allow different predicates to have different declarations.

Among many different prior semantics for rules with aggregations,
there are even different extensions for the same prior semantics for rules with negation, as shown in the right column in Table~\ref{tab-FSCS}.
Unfortunately, 
most of them are defined for limited cases, or add some case-specific definitions.
In particular, simple formal explanations for the disagreements, including among all different extensions for each of WFS and SMS, are completely missing.
We are only aware of comparisons by examples or very restricted cases, 
even for disagreeing semantics by the same authors. %
However, for all such examples and cases we examined, we found that the desired results for them correspond to our semantics under some appropriate declarations for some predicates.
These results are described in\lfcsonly{}\fullonly{ Section~\ref{sec-cmp}}.

\def\tabcap{
    \caption{Founded semantics and constraint semantics for rules with aggregation with different declarations (for all predicates in a program), extending prior semantics for rules with negation, and prior extensions.}}
\ieeeonly{}
\begin{table}[htb]
    \lfcsonly{}
    \centering
    
    \lncsonly{}
    \articleonly{
    \begin{tabular}{@{}p{13ex}@{~}p{10ex}@{\,}|@{~}p{25ex}@{\Hex{-10}}r@{~}||@{~}p{32ex}@{}}}
    \ieeeonly{}
    \acmonly{}
Declarations & Semantics
                          & Extending & Reference & Prior Extensions\\
\hline\hline
certain		 & Founded,   & Stratified & Van Gelder 1986
                                    & %
                                      e.g.,~\cite{mumick1990,ross1992monotonic}\\ %
             & Constraint & (Perfect) %
                                      &~\cite{van86negation}\\
\hline\hline
uncertain, 	 & Founded    & (none found) & 
                                    & (none found) \\
             \cline{2-5}
not complete & Constraint & First-Order Logic &
                                    & e.g., %
                                        \cite{hella2001logics}\\
\hline\hline
uncertain,
             & Founded    & Fitting & Fitting 1985~\cite{fitting85}
                                    & Pelov et al.\,2007~\cite{pelov2007well}\\
complete,    &            & (Kripke-Kleene) & &\\
             \cline{2-5}
not closed   & Constraint & Supported & Apt et al.\,1988~\cite{apt88}
                                    & Pelov et al.\,2007~\cite{pelov2007well}\\
\hline\hline
uncertain,
             & Founded    & WFS & Van Gelder et al.\,1988
                                    & Kemp-Stuckey 1991~\cite{kemp1991semantics}\\
                                      \cline{5-5}
complete,    &            &     &~\cite{van88unfounded,van+91well}
                                    & Van Gelder 1992~\cite{van92agg}\\
                                      \cline{5-5}
closed       &            &     &   & Pelov et al.\,2007~\cite{pelov2007well}\\
             \cline{2-5}
             & Constraint & SMS & Gelfond-Lifschitz 1988
                                    & Kemp-Stuckey 1991~\cite{kemp1991semantics}\\
                                      \cline{5-5}
             &            &     &~\cite{GelLif88}
                                    & Pelov et al.\,2007~\cite{pelov2007well}\\
                                      \cline{5-5}
             &            &     &   & Faber et al.\,2011~\cite{faber2011semantics}\\
                                      \cline{5-5}
             &            &     &   & Gelfond-Zhang\lncsonly{} 2014-2019~\cite{gelfond2019vicious}\\
\hline
    \end{tabular}
    \articleonly{\tabcap}\lncsonly{}
    \label{tab-FSCS}
\end{table}

\newcommand{\myparfirst}[1]{\Vex{0}\noindent {\bf #1.}}

\newcounter{thmcounter}
\newenvironment{mytheorem}{
\refstepcounter{thmcounter}
\noindent\textbf{\textit{Theorem \thethmcounter}}.}{\hfill}

\newcommand{\union}{\cup}

\newcommand{\pgm}{\pi}
\newcommand{\comb}{{\it Combine}}
\newcommand{\addinv}{{\it AddInv}}
\newcommand{\cmpl}{{\it Cmpl}}
\newcommand{\lfp}{{\it LFP}}
\newcommand{\lfpscc}{{\it LFPbySCC}}
\newcommand{\addneg}{{\it AddNeg}}

\newcommand{\founded}{{\it Founded}}
\newcommand{\constraint}{{\it Constraint}}

\newcommand{\dg}{{\it DG}}
\newcommand{\nameneg}{{\it NameNeg}}
\newcommand{\unnameneg}{{\it UnNameNeg}}
\newcommand{\proj}[2]{{\it Proj}(#1,#2)}
\newcommand{\merge}{{\it Merge}}
\newcommand{\mpa}{{\it MergeAtom}}
\newcommand{\selffalse}{{\it SelfFalse}}

\newcommand{\NRB}{{\it NRB}}
\newcommand{\Num}{{\it Num}}

\newcommand{\first}{{\rm first}}
\newcommand{\vdashl}{\vdash_{\rm L}}
\newcommand{\vdashe}{\vdash_{\rm E}}
\newcommand{\dc}{{\it DC}}

\mysec{Language}
\label{sec-lang}

We consider Datalog rules extended with unrestricted negation, disjunction,
quantification, aggregation, and comparison containing aggregation.

\mypar{Domain}
The {\em domain} of a program is the set of values that variables can be instantiated with.  These values are called {\em constants}.  The domain includes the values that appear in the program and a set $\Num$ of numbers.  $\Num$ is a bounded range of numbers determined by a {\em numeric representation bound} $\NRB$ and a {\em numeric representation precision} ${\it NRP}$, i.e., $\Num$ contains all numbers in the range $[-\NRB, \NRB]$ with at most ${\it NRP}$ decimal places.  Numbers with more than ${\it NRP}$ decimal places that appear in the program or arise during evaluation can be rounded to ${\it NRP}$ decimal places\fullonly{ (such rounding can be reported)}, or a higher-precision representation can be used.

This rounding or increasing precision is not shown explicitly in the semantics, because the rule language in this paper\fullonly{ (which includes all major aggregation operations: count, max, min, and sum)} does not include numeric operations that increase the number of decimal places.  We use an ${\it NRP}$ that is at least the maximum number of decimal places in numbers that appear in the program, so all numeric computations are exact\fullonly{, and rounding or increasing precision is not needed}.
\lfcsonly{Our semantics detects and can report cases where an inference is blocked because it involves a value outside the range $[-\NRB, \NRB]$; for details, see the description of range-blocked inference in \cite{LiuSto20RuleAgg-arxiv}.}

\mypar{Datalog rules with unrestricted negation}
We first present a simple core form of rules and then describe additional
constructs that can appear in rules.  The {\it core form} of a rule is the
following, where any \m{P_i} may be preceded with \NOT:
$$Q(X_1, ... ,X_a) ~\IF~ P_1(X_{11}, ... ,X_{1a_1}) ~\AND~ ... ~\AND~ P_h(X_{h1}, ... ,X_{ha_h})$$
\fullonly{Symbols \IF, \AND, and \NOT indicate backward implication, conjunction, and
negation, respectively.
}$Q$ and $P_i$'s are predicates, and each argument \m{X_k} and \m{X_{ij}} is a
constant or a variable.
In\lfcsonly{ arguments of predicates in} examples, we use numbers and quoted strings
for constants and letters for variables.

If \m{h = 0}, then there are no \m{P_i}'s or \m{X_{ij}}'s,
and \lfcsonly{each \m{X_k} must be a constant, in which case} 
\m{Q(X_1,...,X_a)} is called a {\em fact}.
\fullonly{Facts are first transformed so that each \m{X_k} that is a variable is instantiated with all constants in the domain of the program.  So facts have only constants as arguments for the rest of the paper.} Also, for
the rest of the paper, ``rule'' refers only to the case where \m{h \geq 1},
in which case the left side of \fullonly{the backward implication}\lfcsonly{\IF} is called the
{\em conclusion}, the right side is called the {\em body}, and each
conjunct in the body is called a {\em hypothesis}.
We do not require variables in the conclusion to be in the hypotheses; it is not needed because rules are used with variables replaced by constants, and the domain of variables is finite.

\mypar{Disjunction}
In a rule body, hypotheses may be combined using disjunction (\m{\lor}) as well as conjunction.  Conjunction and disjunction may be nested arbitrarily.

 mixed with conjunction and disjunction. 

\mypar{Quantification}
A hypothesis in a rule body can be an existential or universal quantification of the form
\begin{center}
\begin{tabular}[c]{@{}ll@{}}
  $\SOME~ X_1,...,X_a~\co{|}~ B$ & ~~existential quantification\\
  $\EACH~ X_1,...,X_a~\co{|}~ B$ & ~~universal quantification
\end{tabular}
\end{center}
where each $X_i$ is a variable\lfcsonly{that appears} in $B$, 
and $B$ has the same form as a rule body.
Note that this recursive definition allows nested quantifications.
Each quantified variable $X_i$ ranges over the domain of the program.
The quantifications return true iff for some or all, respectively, 
combinations of values of $X_1,...,X_a$, the body $B$ is true.

\fullonly{
To restrict a quantified variable $X_i$ to a particular set of values, 
a user can introduce a predicate $D$ with a fact $D(x)$ for each element $x$ of that set, 
and write an existential quantification 
by adding a conjunct $D(X_i)$ to the body
and write a universal quantification 
by adding a disjunct $\NOT D(X_i)$ to the body.
}

\mypar{Aggregation and comparison}
A {\em set expression} has the form $\{X_1, ... ,X_a: B\}$, where each $X_i$ is a
variable in $B$, and the body $B$ has the same form as a rule body.  The {\it arity} of this set expression is $a$.
The body of each set expression is first rewritten to 
have the same form as the body of a core-form rule,
by introducing auxiliary predicates, e.g., \co{\{Y:~\SOME X | p(X,Y)\}} 
is rewritten to \co{\{Y:~q(Y)\}} together with a new rule \co{q(Y) \IF \SOME X | p(X,Y)}\fullonly{, where \co{q} is a fresh predicate}.
Each auxiliary predicate is declared complete (which is the default, defined below), except that it is declared closed 
if some predicate in the body of the rule defining the auxiliary predicate is declared closed.
\fullonly{This exception preserves any cycle of dependencies among closed predicates.}

An {\em aggregation} has the form ${\it agg}\,S$, where {\it agg} is an
aggregation operator (\co{count}, \co{max}, \co{min}, or \co{sum}), and $S$
is a set expression.  
The aggregation returns the result of applying the
respective {\it agg} operation (cardinality, maximum, minimum, or sum) to the set value of $S$.
\co{max} and \co{min} use the order on numbers, 
extended lexicographically to an order on tuples.
\co{sum} is on numbers, and on tuples whose first components are numbers;
in the latter case, the first components are summed.
Note that \co{count} and \co{sum} applied to the empty set equal 0,
while \co{max} and \co{min} applied to the empty set give an error. %

A hypothesis of a rule may be a {\em comparison} of the form
$${\it agg}\,S \odot k \hspace{2em}\mbox{or}\hspace{2em} 
{\it agg}\,S \odot {\it agg}'\,S'$$
where ${\it agg}\,S$ and ${\it agg}'\,S'$ are aggregations, the comparison operator $\odot$ is an equality ($=$) or inequality ($\ne$, < $\le$, >, $\ge$), and $k$ is a variable or numeric constant or, if the aggregation operator is \co{max} or \co{min}, a tuple of variables or numeric constants.
Comparisons of the second form are first rewritten as two
comparisons of the first form by introducing
a fresh variable.  For example, ${\it agg}\,S \ne {\it agg}'\,S'$
is rewritten as ${\it agg}\,S \ne V \land {\it agg}'\,S' = V$,
and ${\it agg}\,S < {\it agg}'\,S'$ is rewritten as 
${\it agg}\,S < V \land {\it agg}'\,S' \ge V$, where $V$ is a fresh variable.
The latter rewrite uses two inequalities, instead of an inequality and an equality, to increase the cases where occurrences of predicate atoms are positive (defined below).

\lfcsonly{Note that negation applied to comparisons can be eliminated by reversing the comparison operators; for example, the negation of a comparison using $\le$ is a comparison using $>$.}\fullonly{We include a comprehensive set of comparison operators for readability;
it also allows negation applied to comparisons to be eliminated by reversing the comparison operators; 
for example, the negation of a comparison using $<$ is a comparison using $\ge$.}

The key idea here is that the value of a comparison (containing an aggregation) is undefined 
if there is not enough information about the predicates used to
determine the value, or if applying the comparison (containing an aggregation) gives an error, such as a type error.\lfcsonly{ Our principled approach can easily support additional aggregation and comparison functions, e.g., on other data types such as strings.}\fullonly{

Additional aggregation and comparison functions, e.g., %
on other data types such as characters and strings, 
can be supported in %
the same principled way as we support those discussed here.}

\mypar{Programs, atoms, and literals}
A {\em program} $\pgm$ is a set of rules and facts, plus declarations for
predicates, described after dependencies are introduced next.

An {\em atom} of $\pgm$ is either a predicate symbol $P$ in $\pgm$ applied to constants in the domain of $\pgm$ and variables, 
or a comparison formed using predicate symbols in $\pgm$, constants in the domain of $\pgm$, and variables.
These are called {\em predicate atoms} for $P$
and {\em comparison atoms}, respectively.

A {\em literal} of $\pgm$ is either an atom of $\pgm$ or the negation of a
predicate atom of $\pgm$.  These are called {\em positive literals} and
{\em negative literals}, respectively.  A literal that is a predicate atom or its negation is called a {\em predicate literal}.  A literal that is a comparison atom is called a {\em comparison literal}.
Note that negation of a comparison atom is not needed because 
the negation will be eliminated by reversing the comparison operator.

\mypar{Dependency graph}
The dependency graph of a program characterizes dependencies between
predicates induced by the rules, distinguishing positive from non-positive
dependencies.\fullonly{  We define the dependency graph before discussing
declarations for predicates, because the permitted declarations and default
declarations are determined by the dependency graph.}

An occurrence $A$ of a predicate atom in a hypothesis $H$ is 
a {\em positive occurrence} if
(1) $H$ is $A$, which is a positive literal,
(2) $H$ is a quantification, and $A$ is a positive literal in its body,
(3) $H$ is a comparison atom of the form 
\co{count} $S$ \GEQ $k$, \co{count} $S$ \GT $k$,
\co{max} $S$ \GEQ $k$, \co{max} $S$ \GT $k$, 
\co{min} $S$ \LEQ $k$, or \co{min} $S$ \LT $k$,
and $A$ is in a positive literal in the set expression $S$, or
(4) $H$ is a comparison atom of the form 
\co{count} $S$ \LEQ $k$, \co{count} $S$ \LT $k$,
\co{max} $S$ \LEQ $k$, \co{max} $S$ \LT $k$, 
\co{min} $S$ \GEQ $k$, or \co{min} $S$ \GT $k$,
and $A$ is in a negative literal in the set expression $S$.
\fullonly{A predicate atom that is the conclusion of a rule is also a positive occurrence.}
Other occurrences of predicate atoms are {\em non-positive occurrences}.

This definition conservatively
ensures that hypotheses are
monotonic with respect to positive occurrences of predicate atoms, 
i.e., making a positive occurrence of a predicate atom in a hypothesis true
cannot make the hypothesis change from true. %
This definition can be extended so that any occurrence 
\m{A} of a predicate atom in a hypothesis \m{H}
is a {\em positive occurrence} if \m{H} can be determined %
to be monotonic with respective to \m{A}.
For example, if predicate \co{p} holds for only non-negative numbers, then
\co{p(x)} is a positive occurrence in \co{sum \{x:~p(x)\} \GT k}.

The {\em dependency graph} $\dg(\pgm)$ of program $\pgm$ is a directed graph with a node for each predicate of $\pgm$, and an edge from $Q$ to $P$ labeled positive (respectively, non-positive) if a rule whose conclusion contains $Q$ has a hypothesis that contains a positive (respectively, non-positive) occurrence of an atom for $P$.  If there is a path from $Q$ to $P$ in $\dg(\pgm)$, then $Q$ {\em depends on} $P$ in $\pgm$.  If the node for $P$ is in a cycle containing a non-positive edge in $\dg(\pgm)$, then $P$ has {\em circular non-positive dependency} in $\pgm$.

\mypar{Declarations}
A predicate declared {\em certain} means that each assertion of the predicate has a unique true (\T) or false (\F) value.  A predicate declared {\em uncertain} means that each assertion of the predicate has a unique true, false, or undefined (\UD) value. A predicate declared {\em complete} means that all rules with that predicate in the conclusion are given in the program.  A predicate declared {\em closed} means that an assertion of the predicate is set to false, called {\em self-false}, if inferring it to be true using the given rules and facts requires assuming itself to be true.

A predicate must be declared uncertain if it has circular non-positive dependency, or depends on an uncertain predicate; otherwise, it may be declared certain or uncertain and is by default certain. A predicate may be declared complete or not only if it is uncertain, and it is by default complete.  A predicate may be declared closed or not only if it is uncertain and complete, and it is by default not closed.

\lfcsonly{We do not give a syntax for predicate declarations, 
because it is straightforward, and
most examples use default declarations.  However, the language in~\cite{LiuSto20LogicalConstraints-LFCS,LiuSto21LogicalConstraints-JLC} supports such declarations.}
\fullonly{We do not give here a syntax for declarations of predicates to be
certain, complete, closed, or not, because it is straightforward, and
almost all examples use default declarations.  However, Liu and
Stoller~\cite{LiuSto20LogicalConstraints-LFCS,LiuSto21LogicalConstraints-JLC} introduces a language that
supports such declarations and supports the use of both founded semantics 
and constraint semantics.}

\mypar{Notations}
In presenting the semantics, in particular the completion rules, 
we allow negation in the conclusion of rules, and 
we allow hypotheses to be equalities (=) and negated equalities ($\NEQ$) between two variables or a variable and a constant.

\mysec{Formal semantics}
\label{sec-sem}

This section extends the definitions of founded semantics and 
constraint semantics in~\cite{LiuSto18Founded-LFCS,LiuSto20Founded-JLC} 
to handle aggregation and comparison.  
We introduce a new relation, namely, derivability of comparisons, 
and extend most of the foundational definitions,
including the definitions of atom, literal, and positive
occurrence in Section~\ref{sec-lang}, and of
complement, ground instance, truth value of a literal in an interpretation, 
completion rule, naming negation,
unfounded set, and constraint model in this section.
By carefully extending these foundational definitions, 
we are able to avoid explicit changes to the definitions of
other terms and functions built on them, including the definition of
completion function and the definition of the least fixed point at the heart of the
semantics, embodied mainly in the function $\lfpscc$.

\mysubsec{Interpretations and derivability}
\label{sec-derivability}

\myparfirst{Complements and consistency} %
The predicate literals $A$ and $\NOT A$ are {\em complements} of each
other.  The following pairs of comparison literals are complements of each
other: ${\it agg}\,S = k$ and ${\it agg}\,S \ne k$; ${\it agg}\,S \le k$ and
${\it agg}\,S > k$; ${\it agg}\,S \ge k$ and ${\it agg}\,S < k$.  

A set of predicate literals is {\em consistent} if it does not contain a literal and
its complement.

\mypar{Ground instance}
An occurrence of a variable $X$ in a quantification $Q$ is {\em bound} in
$Q$ if $X$ is a variable to the left of the vertical bar in $Q$.
An occurrence of a variable $X$ in a set expression $S$ is {\em bound} if
$X$ is a variable to the left of the colon in $S$.  An occurrence
of a variable in a rule $R$ is {\em free} if it is not bound in a
quantification or set expression in $R$.

A {\em ground atom} or {\em ground literal} is an atom or literal, 
respectively, not containing variables.  
A {\em ground instance} of a rule $R$ in a program $\pgm$ is any rule
obtained from $R$ by expanding universal quantifications into conjunctions
over all constants, 
instantiating existential quantifications
with any constants, and instantiating the remaining free occurrences of
variables with any constants (of course, all free occurrences of the same
variable are replaced with the same constant).  A {\em ground instance} of
a comparison atom $A$ is a comparison atom obtained from $A$ by
instantiating the free occurrences of variables in $A$ with any constants.
A {\em ground instance} of a set expression $\{X_1, ... ,X_a: B\}$ is a
pair $((X_1,...,X_a), B)$ obtained by instantiating all variables in
$X_1,...,X_a$ and $B$ with any constants.

\mypar{Interpretations}
An {\em interpretation} of a program $\pgm$ is a consistent set of ground predicate literals of $\pgm$.  Interpretations are generally 3-valued: a ground predicate literal is {\em true} (i.e., has truth value \T) in interpretation $I$ if it is in $I$, is {\em false} (i.e., has truth value \F) in $I$ if its complement is in $I$, and is {\em undefined} (i.e., has truth value \UD) in $I$ if neither it nor its complement is in $I$.  An interpretation of $\pgm$ is {\em 2-valued} if it contains, for each ground predicate atom $A$ of $\pgm$, either $A$ or its complement.
Interpretations are ordered by set inclusion $\subseteq$.

Let $G(S)$ denote the set of ground instances of set expression $S$.
For a set expression $S$, interpretation $I$, and truth value $t$, let %
$$G(S,I,t) = \{ x \;|\; (x,B) \in G(S) \land \mbox{$B$ has
  truth value $t$ in $I$}\}$$
That is, $G(S,I,t)$ is the set of combinations of constants for which 
the body of set expression $S$ has truth value $t$ in $I$.

\newcommand{\myiff}{\;{\Leftrightarrow}\;}

\mypar{Derivability of comparisons}
\begin{figure*}[t]
\begin{eqnarray*} %
  \pgm,I \vdashl \mco{count}\,S = k &\myiff&
        |G(S,I,\T)| = k \land G(S,I,\UD) = \emptyset\\
  \pgm,I \vdashl \mco{count}\,S > k &\myiff& |G(S,I,\T)| > k\\
  \pgm,I \vdashl \mco{count}\,S < k &\myiff& 
        |G(S,I,\T) \union G(S,I,\UD)| < k\\
 \pgm,I \vdashl \mco{max}\,S = k   &\myiff& 
    k \in G(S,I,\T) \land \EACH~i \in G(S,I,\T) \union G(S,I,\UD)~|~i \le k\\
  \pgm,I \vdashl \mco{max}\,S \ne k &\myiff& 
    k \not\in G(S,I,\T) \union G(S,I,\UD) \lor \SOME~i \in G(S,I,\T)~|~i > k\\
  \pgm,I \vdashl \mco{max}\,S > k   &\myiff& %
    \SOME~i \in G(S,I,\T)~|~i > k\\
  \pgm,I \vdashl \mco{max}\,S < k   &\myiff&
    \SOME~i \in G(S,I,\T) \land \EACH~i \in G(S,I,\T) \union G(S,I,\UD)~|~i < k\\
  \pgm,I \vdashl \mco{sum}\,S = k   &\myiff& 
     {\rm sum}\ G(S,I,\T) = k \land \{\first(i): i \in G(S,I,\UD)\} \subseteq \{0\}\\
  \pgm,I \vdashl \mco{sum}\,S > k   &\myiff& 
     {\rm sum}\ (G(S,I,\T) \union \{i \in G(S,I,\UD): \first(i) < 0 \}) > k\\
  \pgm,I \vdashl \mco{sum}\,S < k   &\myiff& 
     {\rm sum}\ (G(S,I,\T) \union \{i \in G(S,I,\UD): \first(i) > 0 \}) < k   
\end{eqnarray*}
\vspace{-6ex}
\caption{\lfcsonly{\small} Linear-time derivability relation for comparisons.
$\first(i)$ returns the first component of $i$ if $i$ is a tuple, and returns $i$ otherwise.
Biconditionals ($\!\!\mathord{\myiff}\!\!$) for derivability of other comparisons 
are obtained from those given as follows.  
(1)
Biconditionals for deriving comparisons using \co{min} are obtained
from those for \co{max} by replacing \co{max} with \co{min}, 
interchanging \m{\leq} and \m{\geq}, and interchanging $<$ and $>$.
(2) For aggregation operator {\it agg} being \co{count} or \co{sum}, the right side of the biconditional for deriving ${\it agg}\,S \ne k$ is the disjunction of the right sides of the biconditionals for deriving ${\it agg}\,S > k$ and ${\it agg}\,S < k$.
(3)
For each aggregation operator {\it agg}, biconditionals for deriving
${\it agg}\,S \ge k$ and  ${\it agg}\,S \le k$ 
are obtained from the given biconditionals for 
${\it agg}\,S > k$ and ${\it agg}\,S < k$, respectively, 
by replacing $> k$ with $\ge k$ and replacing $< k$ with $\le k$.
}
\label{fig:derivable}
\end{figure*}
Informally, a ground comparison atom ${\it agg}\,S \odot k$ is {\em derivable} in interpretation $I$ of $\pgm$,
denoted $\pgm,I \vdash {\it agg}\,S \odot k$, if the comparison must be true in $I$,
regardless of whether atoms with truth value \UD are true or false.

Precisely, founded semantics uses the {\em linear-time derivability relation} $\vdashl$ defined in Figure~\ref{fig:derivable} based on the aggregation operator and the comparison operator.  It can be computed straightforwardly in linear time in $|G(S,I,\T)|+|G(S,I,\UD)|$.

Derivability for each comparison in Figure~\ref{fig:derivable} has also a condition that the comparison does not give an error. 
It gives an error if
the aggregation gives an error, or if there is a type error, i.e., 
either the aggregation is \co{count} or \co{sum}, or is \co{max} or \co{min} with arity of $S$ being 1, and $k$ is not a number, 
or the aggregation is \co{max} or \co{min} with arity $a$ of $S$ greater than 1, and $k$ is not an $a$-tuple of numbers.
The aggregation gives an error if it is \co{max} or \co{min} and $G(S,I,\T)\union G(S,I,\UD)$ is empty, or if there is a type error, i.e., either it is \co{max} or \co{min} and $G(S,I,\T)$ or $G(S,I,\UD)$ contains either a non-number or a tuple containing a non-number, 
or it is \co{sum} and $S$ has arity 1 and $G(S,I,\T)$ or $G(S,I,\UD)$ contains a non-number,
or it is \co{sum} and $S$ has arity greater than 1 and $G(S,I,\T)$ or $G(S,I,\UD)$ contains a tuple whose first component is not a number.
Comparisons that give errors can easily be detected and reported by checking these conditions.

This definition of derivability is relatively strict about errors, for example, it always makes a comparison give an error if the aggregation in it gives an error.
One can be less strict about errors, for example, a comparison containing \co{max} or \co{min} applied to the empty set and using negated equality could be allowed to hold even if the aggregation in it gives an error, taking the view that an error is not equal to a value or a tuple of values in the domain.  This generally yields more literals that are true or false, rather than undefined.  Choices for error handling could also be specified using declarations.

An alternative to linear-time derivability is {\em exact derivability}, denoted $\vdashe$.  Informally, $\pgm,I \vdashe {\it agg}\,S \odot k$ holds iff (1) ${\it agg}\,S \odot k$ holds in all 2-valued interpretations $I'$ that extend $I$ and satisfy the part of $\pgm$ that $S$ depends on, and (2) there is at least one such interpretation $I'$.  
Exact derivability is based on enumeration of interpretations 
and hence is less appropriate for founded semantics, which is designed to leave such enumeration for constraint semantics.  
Although exact derivability can be more precise in principle, linear-time derivability gives the same result as exact derivability for all examples we found in the literature.

Interpretations provide truth values for comparison literals similarly as for predicate literals.  Let $\dc(\pgm,I)$ be the set of comparisons derivable for program $\pgm$ and interpretation $I$.  
A comparison literal $A$ for $\pgm$ is {\em true} in $I$ if it is in $\dc(\pgm, I)$, is {\em false} in $I$ if its complement is in $\dc(\pgm,I)$, and is {\em undefined} in $I$ otherwise. 

\mypar{Models}
An interpretation $I$ of a program $\pgm$ is a {\em model} of $\pgm$ if it (1)
contains all facts in $\pgm$, and (2) satisfies all rules of $\pgm$,
interpreted as formulas in 3-valued logic \cite{fitting85} (i.e., for
each ground instance of each rule, if the body is true in $I$, then so is
the conclusion).

\mypar{One-step derivability}
The {\em one-step derivability} function $T_\pgm$ for program $\pgm$
performs one step of inference using rules of $\pgm$.
Formally, $A\in T_\pgm(I)$ iff (1)
$A$ is a fact of $\pgm$, or (2) there is a ground instance $R$ of a rule of
$\pgm$ with conclusion $A$ such that the body of $R$ is true in
interpretation $I$.

\mysubsec{Founded semantics without closed declarations}
\label{sec-founded}

We first define a version of founded semantics, denoted $\founded_0$, that
ignores declarations that predicates are closed.  
We then extend the definition to handle those
declarations. %
Intuitively, the {\em founded model} of a program $\pgm$ ignoring closed-predicate declarations, denoted
$\founded_0(\pgm)$, is the least set of literals that are given as facts or
can be inferred by repeatedly applying the rules. 
Formally, we define
\lncsonly{$$\founded_0(\pgm) = \unnameneg(\lfpscc(\nameneg(\cmpl(\pgm)))),$$}%
\articleonly{$$\founded_0(\pgm) = \unnameneg(\lfpscc(\nameneg(\cmpl(\pgm)))),$$}%
\acmonly{$$\founded_0(\pgm) = \unnameneg(\lfpscc(\nameneg(\cmpl(\pgm)))),$$}%
\ijcaionly{$$\resizebox{\linewidth}{!}{$%
\founded_0(\pgm) = \unnameneg(\lfpscc(\nameneg(\cmpl(\pgm)))),$}$$}%
\ieeeonly{$$\resizebox{\linewidth}{!}{$%
\founded_0(\pgm) = \unnameneg(\lfpscc(\nameneg(\cmpl(\pgm)))),$}$$}%
where functions $\cmpl$, $\nameneg$, $\lfpscc$, and $\unnameneg$ are defined as follows.

\mypar{Completion}
The completion function $\cmpl(\pgm)$ returns the {\em completed program} of
$\pgm$.  Formally, $\cmpl(\pgm)=\addinv(\comb(\pgm))$, where
$\comb$ and $\addinv$ are defined as follows.

The function $\comb(\pgm)$ returns the program obtained from $\pgm$ 
by replacing the facts and rules defining each uncertain complete predicate $Q$ with a single {\em combined rule} for $Q$, defined as follows.  First, transform the facts and rules defining $Q$ so they all have the same conclusion $Q(V_1,...,V_a)$, by replacing each fact or rule $Q(X_1,...,X_a) ~\IF~ B$ with 
$$Q(V_1,...,V_a) ~\IF~ (\SOME~Y_1,...,Y_k ~|~ V_1=X_1 \land \,...\, \land V_a=X_a \land B)$$
where $V_1,...,V_a$ are fresh variables (i.e., not occurring in 
any given rule defining $Q$), and $Y_1,...,Y_k$ are all variables occurring free in 
the original rule $Q(X_1,...,X_a) ~\IF~ B$. 
Then, combine the resulting rules for $Q$ into a single rule 
defining $Q$ whose body is the disjunction of the bodies of those rules. 
This combined rule for $Q$ is logically equivalent to the original facts and rules for $Q$.

The function $\addinv(\pgm)$ returns the program obtained from $\pgm$ by
adding, for each uncertain complete predicate $Q$, a {\em completion rule}
that derives negative literals for $Q$.  The completion rule for $Q$ is
obtained from the inverse of the combined rule defining $Q$ (recall that
the inverse of $A~\IF~B$ is $\NOT A~\IF~\NOT B$), by (1) putting the body
of the rule in negation normal form, i.e., using laws of predicate logic
to move negation inwards and eliminate double negations, and (2) eliminating
negation applied to comparison atoms by reversing the comparison operators.  As
a result, in completion rules, negation is applied only to predicate atoms. 

Similar completion rules but without aggregation are used in Clark's completion~\cite{clark78} and Fitting semantics~\cite{fitting85}.
 
\mypar{Least fixed point}
The least fixed point is preceded and followed by functions that introduce and remove, respectively, new predicates representing the negations of the original predicates.

The function $\nameneg(\pgm)$ returns the program obtained from $\pgm$ by
replacing, except where $P(X_1,...,X_a)$ is a positive occurrence,
$\NOT P(X_1,...,X_a)$ with $\mco{n.}P(X_1,...,X_a)$, and
$P(X_1,...,X_a)$ not \mbox{after} $\NOT$ with $\NOT\, \mco{n.}P(X_1,...,X_a)$.
The new predicate $\mco{n.}P$ represents
the negation of predicate $P$.
Since $P(X_1,...,X_a)$ and $\NOT P(X_1,...,X_a)$ are complements of each other, we now also 
define $P(X_1,...,X_a)$ and $\mco{n.}P(X_1,...,X_a)$ to be complements of each other.

Note that $\mco{n.}P(X_1,...,X_a)$ is introduced to make the one-step
derivability function explicitly monotonic, while maintaining consistency.
We replace $\NOT P(X_1,...,X_a)$ for any conclusion and any negative
occurrence of $P(X_1,...,X_a)$\fullonly{ in a hypothesis} (where negative occurrence is defined
symmetrically as positive occurrence)\fullonly{ with $\mco{n.}P(X_1,...,X_a)$} to allow negative conclusions to be
derived and used as facts.
We replace any negative occurrence of $P(X_1,...,X_a)$ not after 
$\NOT$ with $\NOT\, \mco{n.}P(X_1,...,X_a)$ also
to use these facts.
Other occurrences\fullonly{ of $\NOT P(X_1,...,X_a)$ and $P(X_1,...,X_a)$ in hypotheses}, 
if any due to positive (and
negative) occurrence being conservative, can be either replaced or left,
with the result still being a model, because all derivation and use of
$\mco{n.}P(X_1,...,X_a)$ and $P(X_1,...,X_a)$ follow the one-step
derivability.
We have not seen any example that needs \lfcsonly{this}\fullonly{the following}, but one might obtain a more precise model, i.e., more atoms that are true or false, by trying all combinations of replacing and leaving.
It is an open question whether some combination leads to a unique most precise model.

The function $\lfpscc(\pgm)$ uses a least fixed point to infer facts for
each strongly connected component (SCC) in the dependency graph of $\pgm$,
as follows.  Let $C_1,...,C_n$ be a list of the SCCs in dependency order,
so predicates in earlier SCCs do not depend on predicates in later ones; it is easy to show that any
linearization of the dependency order leads to the same result for
$\lfpscc$.
The {\em projection} of a program $\pgm$ onto 
an SCC $C$, denoted $\proj{\pgm}{C}$, contains all facts of $\pgm$ whose
predicates are in $C$ and all rules of $\pgm$ whose conclusions contain
predicates in $C$.

Define $\lfpscc(\pgm) = I_n$, where $I_0= \emptyset$  and $I_i =
\addneg(\acmonly{$ $}\lfp(T_{\proj{\pgm}{C_i} \union I_{i-1}}),\articleonly{ }\lncsonly{$ $}C_i)$ for $i \in
1..n$.  $\lfp$ is the least fixed point operator.  
$\addneg(I, C)$ returns the interpretation obtained from interpretation $I$
by adding {\em completion facts} for the certain predicates in $C$ to $I$;
specifically, for each certain predicate $P$ in $C$, and each combination
of values $v_1,...,v_a$ of arguments of $P$, if $I$ does not contain
$P(v_1,...,v_a)$, then add $\co{n.}P(v_1,...,v_a)$.\fullonly{

} The least fixed point is well-defined, because the one-step derivability
function $T_{\proj{\pgm}{C_i} \union I_{i-1}}$ is monotonic with respect 
to $\subseteq$, i.e., for all interpretations $J$ and $J'$, 
$T_{\proj{\pgm}{C_i} \union I_{i-1}}(J) \subseteq 
T_{\proj{\pgm}{C_i} \union I_{i-1}}(J')$  whenever $J\subseteq J'$\lfcsonly{; the proof is straightforward \cite{LiuSto20RuleAgg-arxiv}.}\fullonly{.
To establish this, we show monotonicity of each part of the definition of $T_\pgm$.
Part (1) adds a fixed set of facts and hence is trivially monotonic.  
Part (2) is monotonic because derivability of comparisons is 
monotonic with respect to $\subseteq$ 
(i.e., $\dc(\pgm,J) \subseteq \dc(\pgm,J')$ whenever $J\subseteq J'$),
and because every rule that can ``fire'' in $J$ 
(i.e., its hypotheses are true) can also fire in $J'$.}

The function $\unnameneg(I)$ returns the interpretation obtained from
interpretation $I$ by replacing each atom $\co{n.}P(X_1,...,X_a)$ with
$\NOT P(X_1,...,X_a)$.

\mysubsec{Founded semantics with closed declarations}
\label{sec-founded-closed}

Informally, when an uncertain complete predicate is declared closed, an
atom $A$ of the predicate is false in an interpretation $I$ for a program
$\pgm$, called {\em self-false} in $I$, if every ground instance of a rule
that concludes $A$
has a hypothesis that is false in $I$ or, recursively, is self-false in
$I$.  To simplify the formalization, 
we first transform ground instances of rules to eliminate disjunction,
by putting the body of each ground instance $R$ of a rule into disjunctive
normal form (DNF) %
and then replacing $R$ with multiple rules, one per disjunct of the DNF.

A set $U$ of ground predicate atoms for closed predicates is an {\em unfounded
  set} of $\pgm$ with respect to an interpretation $I$ of $\pgm$ iff $U$ is 
disjoint from $I$ and, for each atom $A$ in $U$, and each ground instance $R$ of a rule of $\pgm$
with conclusion $A$, 
\begin{enumerate}
\setlength{\itemsep}{0ex}

\item[(1)] some hypothesis of $R$ is false in $I$,
\item[(2)] some
positive predicate hypothesis of $R$ is in $U$, or
\item[(3)] some comparison hypothesis $H$ of $R$ is false when all atoms in $U$
are false, i.e., $\pgm, I \union \neg\cdot U \vdashl \NOT H$, 

\end{enumerate}
where, for a set $S$ of positive literals,
$\neg \cdot S = \{ \NOT P(c_1,...,c_a) \,|\, P(c_1,...,c_a) \in S\}$,
called the {\em element-wise negation} of \m{S}, and where $\NOT H$ is
implicitly simplified to eliminate negation applied to \m{H} by changing
the comparison operator in \m{H}.

Note that this definition differs from the standard definition of unfounded set~\cite{van+91well}
in that we restricted the unfounded set to atoms for closed predicates, 
added clause (3), and added the disjointness condition.
Because a comparison hypothesis depends non-conjunctively on 
the truth value\fullonly{s} of multiple literals for predicates used in the aggregation, and these literals may be spread across $I$ and $U$, 
clause (3) checks whether $H$ is false when all atoms in $U$ are set to false.
The explicit disjointness condition is not needed in WFS or founded semantics without aggregation, because one can prove in those settings that unfounded sets are disjoint from interpretations that arise in the semantics (e.g., see~\cite[Lemma 3.4]{van+91well}).  The disjointness condition is needed here to ensure that the interpretation $I\union\neg \cdot U$ in clause (3) is consistent and hence the meaning of the clause is well-defined.

The definition of unfounded set $U$ ensures that extending $I$ to make all
atoms in $U$ false is consistent with $\pgm$, in the sense that no atom in
$U$ can be inferred to be true in the extended interpretation.
We define
$\selffalse_\pgm(I)$, the set of {\it self-false atoms} of $\pgm$ with respect to
interpretation $I$, to be the greatest unfounded set of $\pgm$ with respect to
$I$.
Note that this set is empty when no predicate is declared closed. 

The founded semantics is defined by repeatedly computing the semantics
given by $\founded_0$ (founded semantics without closed declarations) and
then setting self-false atoms to false, until a least fixed point is
reached.  Formally, the founded semantics is $\founded(\pgm)=\lfp(F_\pgm)$,
where
$F_\pgm(I) = \founded_0(\pgm \union I) \union \neg \cdot
\selffalse_\pgm(\founded_0(\pgm \union I))$.

\mysubsec{Constraint semantics}
\label{sec-constraint-sem}

Constraint semantics is a set of 2-valued models based on founded semantics.  A {\em constraint model} $M$ of a program $\pgm$ is a 2-valued interpretation of $\pgm$ such that (1) $\founded(\pgm) \subseteq M$, (2) $M$ is a model of $\cmpl(\pgm)$, and (3) if there are closed predicates, there is no non-empty subset $S$ of $M\setminus\founded(\pgm)$ such that $S$ contains only positive literals for closed predicates and $S = \selffalse_\pgm(M\setminus S)$.
Condition (3) says that $M$ should not contain a set $S$ of positive literals for closed predicates that are not required to be true by the founded semantics and are self-false with respect to the rest of $M$.

We also require that an interpretation that leads to an error in a comparison be not a constraint model.
Precisely, we require that for interpretation $M$ to be a constraint model, no ground instance of a rule of $\pgm$ contains a comparison that gives an error in $M$.  Errors are defined the same as in Section \ref{sec-derivability}, but note that $G(S,I,\UD)$ is empty here.
This definition of constraint models could\fullonly{ also} be made less strict about errors as in Section~\ref{sec-derivability}.

Note that condition (3) differs from the corresponding condition in constraint semantics without aggregation~\cite{LiuSto18Founded-LFCS,LiuSto20Founded-JLC}, which is $\neg \cdot \selffalse(M) \subseteq M$.  The change is needed because of the new disjointness condition for unfounded sets.  With the new disjointness condition, for any 2-valued interpretation $M$, $\selffalse(M)$ must be empty, and hence $\neg \cdot \selffalse(M) \subseteq M$ is vacuously true.

We define $\constraint(\pgm)$ to be the set of constraint models of $\pgm$.
Constraint models can be computed by iterating over interpretations $M$ that are supersets of $\founded(\pgm)$, thus satisfying condition (1), and
then checking whether conditions (2) and (3) are satisfied.

\fullonly{
\mysubsec{Range-blocked inference}
\label{sec-range}

Use of \co{sum} or \co{count} can generate values outside the representable range $[-\NRB, \NRB]$.
Our semantics only considers predicate argument values in the domain, and thus does not infer facts with an argument value outside the representable range.
We call inferences that would derive such facts ``range-blocked inferences''.

Formally, program $\pgm$ has a {\em range-blocked inference} in interpretation $I$ if there is a rule $R$ of $\pgm$ such that: (1) $R$ has a hypothesis of the form ${\it agg}\,S = k$, (2) $k$ is a variable that also occurs in the conclusion of $R$,
and (3) there is a ground instance $R'$ of $R$, containing a ground instance ${\it agg}\,S' = k'$ of ${\it agg}\,S = k$, such that (i) all hypotheses of $R'$ other than ${\it agg}\,S' = k'$ are true in $I$ and (ii) either $\pgm, I \models {\it agg}\,S' > \NRB$ or $\pgm, I \models {\it agg}\,S' < -\NRB$.  
We can detect and report range-blocked inferences by checking the conditions in this definition.

This definition is designed to be permissive and allows inferences involving intermediate values (i.e., values not in the inferred facts) outside the representable range.  For example, it allows inference using the rule \co{q \IF sum \{x:~p(x)\} \GT 0} even when the sum is outside the representable range.  The definition can easily be extended to block such inferences as well; this may better reflect implementation behavior, because implementations need to compute intermediate values.
}

\lfcsonly{
\mysubsec{Properties of the semantics}
\label{sec-props}

We briefly state several important properties of the semantics; detailed statements and proofs are in \cite{LiuSto20RuleAgg-arxiv}. 
(1) {\it Consistency:} The founded model and constraint models of a program $\pgm$ are consistent. 
(2) {\it Correctness:} The founded model of a program $\pgm$ is a model of $\pgm$ and $\cmpl(\pgm)$.  The constraint models of $\pgm$ are 2-valued models of $\pgm$ and $\cmpl(\pgm)$.  
(3) {\it Same SCC, same certainty:} All predicates in an SCC have the same certainty. 
(4) {\it Higher-order programming:} Founded semantics and constraint semantics are preserved by a transformation that facilitates higher-order programming by replacing a set $S$ of compatible predicates with a single predicate \co{holds} whose first argument is the name of one of those predicates.  
(5) {\it Equivalent declarations:} Changing predicate declarations from uncertain, complete, and closed to certain when allowed, or vice versa, preserves founded and constraint semantics.
}

\fullonly{
\mysec{Properties of the semantics}
\label{sec-props}

\newenvironment{myproof}[1]{\noindent{\bf Proof#1. }}{\hfill$\blacksquare$\Vex{1}}

\myparfirst{Consistency and correctness}

\begin{mytheorem}
  The founded model and constraint models of a program $\pgm$ are consistent.
  \label{thm:consistent}
\end{mytheorem}

\begin{myproof}{}
See Appendix~\ref{sec-proofs}.
\end{myproof}

\begin{mytheorem}
  The founded model of a program $\pgm$ is a model of $\pgm$ and $\cmpl(\pgm)$.  The constraint models of $\pgm$ are 2-valued models of $\pgm$ and $\cmpl(\pgm)$.
  \label{thm:model}
\end{mytheorem}

\begin{myproof}{}
See Appendix~\ref{sec-proofs}.
\end{myproof}

\mypar{Same SCC, same certainty}
All predicates in an SCC have the same certainty.
The proof is the same as for Theorem 4 in \cite{LiuSto20Founded-JLC}.

\begin{mytheorem}
  For every program, for every SCC $C$ in its dependence graph, all predicates in $C$ are certain, or all of them are uncertain.
  \label{thm:scc-certainty}
\end{mytheorem}

\mypar{Equivalent declarations}
Changing predicate declarations from uncertain, complete, and
closed to certain when allowed, or vice versa, preserves founded and constraint semantics. The same SCC, same certainty property above implies that this change needs to be made for all predicates in an SCC.  This property is formally stated in the following theorem.  Several modifications to the proof of the corresponding theorem in \cite{LiuSto20Founded-JLC} are needed.

When predicates are declared certain, founded semantics and constraint semantics are the same and can be computed much more efficiently than the worst case, by simple least fixed-point computations, without using completion rules, finding self-false atoms, or doing constraint solving.

\begin{mytheorem}
Let $\pgm$ be a program.  Let $C$ be an SCC in its dependency graph containing only predicates that are uncertain, complete, and closed.  Let $\pgm'$ be a program identical to $\pgm$ except that all predicates in $C$ are declared certain.  Note that, for the declarations in both programs to be allowed, predicates in all SCCs that follow $C$ in dependency order must be uncertain, predicates in all SCCs that precede $C$ in dependency order must be certain, and predicates in $C$ must not have circular non-positive dependency.
Then $\founded(\pgm)=\founded(\pgm')$ and $\constraint(\pgm)=\constraint(\pgm')$.
\label{thm:uncertain-certain}
\end{mytheorem}

\begin{myproof}{}
See Appendix~\ref{sec-proofs}.
\end{myproof}

\mypar{Higher-order programming}
Founded semantics and constraint semantics are preserved by a transformation that facilitates higher-order programming by replacing a set $S$ of compatible predicates with a single predicate \co{holds} whose first argument is the name of one of those predicates.  For example, if \co{win} is in $S$, then \co{win(x)} is replaced with \co{holds('win',x)}.  The definitions of compatible predicates and of the program transformation %
that replaces predicates in $S$ with \co{holds} are the same as in \cite{LiuSto20Founded-JLC}, with the clarification that the auxiliary function $\mpa_S(A)$, which replaces predicates in $S$ with \co{holds} in an atom $A$, is applied only to predicate atoms.
The proof is the same as for Theorem 5 in \cite{LiuSto20Founded-JLC}.

\begin{mytheorem}
  Let $S$ be a set of compatible predicates of program $\pgm$.  Then  $\merge_S(\pgm)$ and $\pgm$ have the same founded semantics, in the sense that $\founded(\merge_S(\pgm)) = \mpa_S(\founded(\pgm))$.  $\merge_S(\pgm)$ and $\pgm$ also have the same constraint semantics, in the sense that $\constraint(\merge_S(\pgm))$ ${}= \mpa_S(\constraint(\pgm))$.
  \label{thm:pred-merge}
\end{mytheorem}
} %

\begin{fullversion}
\mysec{Discussion} %
\label{sec-cmp}

This section discusses and compares with major prior semantics for recursive rules with aggregation.
In these comparisons, including the given results for examples, default declarations for the predicates 
are used in our semantics except where stated otherwise.  In particular, %
our semantics is able
to match all desired results from prior semantics.

\mypar{Stratified and monotonic programs}
Many prior works, especially in deductive databases,
consider only stratified programs, which prohibit recursion through aggregation or negation, or monotonic programs, e.g.,~\cite{mumick1990,ross1992monotonic,shkapsky2015optimizing,yang2017scaling,zaniolo2017fixpoint,zaniolo19mono}.
These works vary in the details of the rule language and the programs recognized as stratified or monotonic. 
In our semantics, all these programs fall in the class of programs where all predicates can be declared certain, and whose semantics are fully determined by least fixed-point computations.

\mypar{First-order logic}
Several prior works extend first-order logic with various kinds of aggregations.  For example, Hella et al.~\cite{hella1999logics,hella2001logics} extend first-order logic with aggregation operators similar to those in database query languages, and analyze its expressiveness.  Programs in our language can be regarded as sets of formulas in such logics, with $\IF$ representing reverse implication, and with aggregations translated to accommodate syntactic differences between the languages.  If all predicates in the program are uncertain and not complete, constraint semantics of the program matches the semantics of those formulas, i.e., the constraint models of the program are exactly the models of those formulas.  Theorem 7 in \cite{LiuSto20Founded-JLC} establishes this result without aggregations.  It is easy to see that the result still holds when aggregations are considered, because our semantics for aggregations in 2-valued models matches the standard semantics for aggregations used in such logics.

\mypar{Kemp-Stuckey 1991}
Kemp and Stuckey~\cite{kemp1991semantics} is one of the earliest
comprehensive studies, improving over a number of previous works that
handled limited classes of programs with aggregations.  
They extend WFS and SMS to logic programs with aggregations and study 
previously defined classes of programs with aggregations under several notions of
stratification as well as properties such as monotonicity.  

Their extension of WFS
requires that a set be fully defined before aggregation can be evaluated on it;
thus, if the predicate used in an aggregation is $\UD$ for any argument, 
the truth value of the hypothesis containing that aggregation is also $\UD$.  This is much less precise 
than founded semantics, in the sense that it leaves many more atoms with truth value
$\UD$.  For example, for the company control problem~\cite{mumick1990}, 
they point out that their extension of WFS does not infer the expected results,
while founded semantics and constraint semantics do, as shown in Section~\ref{sec:company-control}.

Another way in which founded semantics is more
precise is that their extension of WFS uses the standard definition of unfounded set, 
while our semantics extends that definition to handle aggregation precisely,
allowing larger unfounded sets and thus more self-false atoms, yielding fewer undefined values.  
For example, for the correlated counts problem discussed in Section \ref{sec:correlated-counts}, 
if \co{p} is declared uncertain, complete, and closed, instead of the default certain,
this additional clause is needed to conclude that \m{\{\mco{p(2)},\mco{p(3)}\}} is an unfounded set.

In extending SMS to handle aggregations, they treat literals containing aggregation as negative literals 
when computing the program reduct.  However, they found that this may produce un-intuitive non-minimal stable models.  For example, for the company control problem, it allows undesired stable
models~\cite[Section 6]{kemp1991semantics},
while founded semantics and constraint semantics yield exactly the desired model, as shown in Section~\ref{sec:company-control}.

\mypar{Van Gelder 1992}
Van Gelder~\cite{van92agg} presents an early approach in which aggregations
are defined using ordinary rules, rather than introduced as new primitives,
in a language with 3-valued semantics.  It illustrates the approach using
examples involving min, max, subset, and sum, with the rules defining the
aggregations customized in some cases to the problem at hand, with the comment 
``{\it ad hoc} methods of analysis seem to be necessary.'' 

The paper
shows that the desired results are obtained for several non-trivial
examples but not for some others.  For example, for the company control problem in Section \ref{sec:company-control}, Van Gelder points out that his semantics sometimes leaves atoms for the \co{controls} predicate undefined, even though they can be determined to be true or false using the given \co{ownsStk} facts~\cite[Example 6.1]{van92agg}; our semantics infers all of the expected \co{controls} relations.

Unfortunately, it is hard to characterize the programs for which their
approach gives desired results because of their use of ad hoc methods.  
In contrast, our work handles a clearly
defined language of programs with aggregations, allows specification of
different assumptions, and supports both 2-valued and 3-valued semantics.
Also, our work allows rules with disjunction and quantifiers.  These are
not considered in~\cite{van92agg}.

\mypar{Pelov et al. 2007}
Pelov, Denecker, and Bruynooghe \cite{pelov2007well} present a
framework in which semantics of logic programs are 
obtained by approximating the fixed points of an immediate consequence operator.  
They define the {\em ultimate approximation} $U^{\it aggr}$ of this operator 
extended to handle aggregations, and define the ultimate Fitting (Kripke-Kleene), 
ultimate WFS, and ultimate SMS of programs with aggregations as appropriate kinds 
of fixed points of  $U^{\it aggr}$.

Their ultimate semantics are intended to be most precise
but have very high computational complexity.  Therefore, Pelov et al.\ 
also introduce more tractable semantics.  
They define extensions $\Phi^{\it agg}$ of Fitting's 3-valued 
approximation of the immediate consequence operator $\Phi$ to handle aggregations,
and obtain extensions of %
Fitting semantics, WFS, and SMS as appropriate kinds of fixed points 
of $\Phi^{\it agg}$.  They consider two versions of $\Phi^{\it agg}$, based
on ``trivial'' and ``ultimate'' 3-valued approximations of the semantics of aggregations.

To see the precision of their ultimate Fitting semantics and ultimate WFS, consider the program 
\begin{code}
  p(1)   p(-1)
  q \IF \NOT q
  c \IF sum \{x: p(x) \AND q\} \GEQ 0
\end{code}
In both their ultimate Fitting semantics and ultimate WFS, \co{c} is undefined.
To see this, first note that \co{q} is undefined, so possible values of the set being summed range from $\emptyset$ (in interpretations where \co{q} is false) to $\{\mco{-1},\mco{1}\}$ (in interpretations where \co{q} is true).  Their framework uses intervals in the subset lattice, so the comparison in the third rule is evaluated for all sets in the interval from $\emptyset$ to $\{\mco{-1},\mco{1}\}$ in the subset lattice, and the comparison is true for some of them (e.g., $\emptyset$) and false for others ($\{\mco{-1}\}$), so \co{c} is left undefined.  Founded semantics 
matches that result, because the comparison in the third rule is not derivable.
Founded semantics using exact derivability would give a more precise result: the comparison is evaluated only for $\emptyset$ and $\{\mco{-1},\mco{1}\}$ as possible values of the set expression, and is true for both, so \co{c} is true, as one would expect for a most precise semantics.

To explore assumptions underlying their extension of SMS, consider the following program used by Faber et al.~\cite[Example 5.2]{faber2011semantics} to compare with other semantics including Pelov et al.'s:
\begin{code}
  p(1) \IF \NOT p(-1)
  p(-1) \IF p(1)
  p(1) \IF sum \{x: p(x)\} \GEQ 0
\end{code}
In Faber et al.'s semantics, this program has one model, $\{\mco{p(1)}, \mco{p(-1)}\}$.  Our constraint semantics (with default declarations) matches Faber et al.'s semantics.
In Pelov et al.'s semantics, this program has no stable models.
Our constraint semantics, when \co{p} is declared closed (which is not the default), matches Pelov et al.'s semantics.

\mypar{Faber et al.~2011}
Faber, Pfeifer, and Leone \cite{faber2011semantics} define an answer set 
semantics (analogous to SMS) for disjunctive programs with aggregations, 
using a generalization of the Gelfond-Lifschitz transformation.  They also analyze 
the computational complexity of the language for several different 
restrictions on the allowed aggregations.

To explain their semantics that differs from other prior semantics, 
consider the following two programs used by Gelfond and Zhang \cite[Example 2]{gelfond2019vicious} to help motivate a different semantics.
The first program, called $P_2$, is
\begin{code}
  p(1) \IF count \{x: p(x))\} \GEQ 0
\end{code}
The second, called $P_3$, is a variant of $P_2$ logically expected to have 
the same meaning:
\begin{code}
  p(1) \IF count \{x: p(x))\} = y, y \GEQ 0
\end{code}
In Faber et al.'s semantics,
$P_2$ has one model, \{\co{p(1)}\}, as in semantics for simple monotonic programs.
Our semantics matches this: \co{p} is certain by default, because it has only a positive dependence on itself, 
and the founded semantics and constraint semantics are the same, giving the same one model.

In Faber et al.'s semantics, $P_3$ has no model, which may be surprising.  
Founded semantics provides an explanation: \co{p} is uncertain and complete by default, 
and \co{p(1)} is undefined.
Constraint semantics gives the same one model as for $P_2$, 
matching the logically expected equivalence with $P_2$.

\mypar{Gelfond-Zhang 2019}
Gelfond and
Zhang~\cite{gelfond2002representing,gelfond2014vicious,zhang2016characterization,gelfond2017vicious,gelfond2018vicious,gelfond2019vicious}
study the challenges and solutions for aggregation in recursion
extensively, in an effort to establish the desired semantics for
aggregation that corresponds to SMS, a set of 2-valued models.  This
resulted in changes from earlier semantics by
Gelfond~\cite{gelfond2002representing}, essentially to capture an implicit
closed-world assumption.  Their most recent work~\cite{gelfond2019vicious}
introduces a new semantics, 
with programs $P_2$ and $P_3$ above as motivating examples.

They note that many prior semantics unintuitively give different results 
for programs $P_2$ and $P_3$ above.  
They resolve this problem by introducing a semantics 
in which both programs \cite[Example 3]{gelfond2019vicious} have no model.
Having no model for $P_2$ contradicts all other semantics we have seen that support aggregation in recursion, 
including semantics for even monotonic programs supported in deductive databases.
To make $P_2$ and $P_3$ logically equivalent, 
the desired way is to make $P_3$ have the same model \{\co{p(1)}\} as $P_2$.
Our constraint semantics achieves that, as described above.

\mypar{Precision, range, and errors}
Computing aggregations require operations on numbers, which incur the
issues with precision, bound, and errors with numeric computations in
reality.
These issues are not explicitly addressed systematically, or not discussed at
all, in prior work.

We describe the choices and the handling of these issues in our language
and semantics: precision and range under Domain in Section~\ref{sec-lang},
Range-blocked inference in Section~\ref{sec-range}, and errors under
Derivability of comparisons in Section~\ref{sec-derivability} and
interpretations of comparisons in Constraint semantics in
Section~\ref{sec-constraint-sem}.

In general, choices for handling precision, range, errors could also be
specified using declarations.  We do not explicitly name these choices, because
they are well-known in numeric computations in practice, and they are
orthogonal to the declarations about predicates in our language and
semantics.

\end{fullversion}

\lfcsonly{}

\begin{fullversion}
\mysec{Additional examples and applications}
\label{sec-examp}

Many small examples similar to the example in Section~\ref{sec-prob} have
been discussed extensively in the literature.  The most recent
work~\cite{gelfond2019vicious} is most comprehensive in discussing 28
examples; we discuss %
their Examples 1 and 28 to show the range of difficulties they deal with,
Example 15 that resorts to a subset relation, and Example 25 that spans the
most discussion.
We then discuss the well-known challenging company control problem~\cite{mumick1990}
and two even more challenging game problems that generalize the well-known win-not-win game~\cite{LiuSto18Founded-LFCS,LiuSto20Founded-JLC}.
In all cases, founded semantics and constraint semantics are simple and give the desired results.

\mysubsec{Classes needing teaching assistants}

This is Example 1 in~\cite{gelfond2019vicious}.
It considers a complete list of students enrolled in a class \co{c},
represented by a collection of facts:
\begin{code}
  enrolled('c','mike')  enrolled('c','john')  ...
\end{code}
It defines a relation \co{need\_ta(c)} that holds iff class \co{c} needs a
teaching assistant, i.e., the number of students enrolled in the class
is greater than 20, and it gives a second rule for its negation, as
follows:
\begin{code}
  need_ta(c) \IF count \{x: enrolled(c,x)\} \GT 20
  n_need_ta(c) \IF \NOT need_ta(c)
\end{code}

Because \co{enrolled} is certain from the list being complete, and there is
no aggregation in recursion,
\co{need\_ta} is certain by default, and so is \co{n\_need\_ta}.  Thus 
founded semantics and constraint semantics are the same and are straightforward to compute.  First,
\co{need\_ta} is inferred %
by just doing the counting for each \co{c} and then checking if
the count is greater than 20.
Then \co{n\_need\_ta} is computed, simply concluding true for classes for
which \co{need\_ta} is false.

\mysubsec{Graduation requirements---directly using universal quantification} 

This is Example 15 in~\cite{gelfond2019vicious}.  It considers a knowledge
base containing two complete lists of facts, for two relations \co{taken}
and \co{required}:
\begin{code}
  taken('mike','cs1')  taken('mike','cs2')  
  taken('john','cs2')
  required('cs1')  required('cs2')
\end{code}
It introduces a subset relation to define a new relation\acmonly{}
\co{ready\_to\_graduate(s)} that holds if student \co{s} has taken all the
required classes:
\begin{code}
  ready_to_graduate(s) \ijcaionly{}\ieeeonly{}\IF \{c: required(c)\} \SUBSET \{c: taken(s,c)\}
\end{code}
The problem description in~\cite[Example 15]{gelfond2019vicious} says that
using the subset relation ``avoids a more complex problem of introducing
universal quantifiers and some kind of implication in the rules of the
language''.

With our language, the rule can be written directly using universal
quantification and implication, where \co{P\IMPLIES{\,}Q} can be trivially
rewritten as \co{\NOT{\,}P\,\OR{\,}Q}, yielding:
\begin{code}
  ready_to_graduate(s) \ieeeonly{}\IF \EACH c | \NOT required(c) \OR taken(s,c)
\end{code}

Because \co{taken} and \co{required} as given are certain, and there is no
negation or aggregation in recursion, \co{ready\_to\_graduate} is certain
by default and can be computed simply as a least fixed point.  This yields
the same result for founded semantics and constraint semantics:
\co{ready\_to\_graduate('mike')} is true and
\co{ready\_to\_graduate('john')} is false.

Example 15 in~\cite{gelfond2019vicious} also discusses other assumptions and rules.  They are
either non-issues or straightforward to handle in our language.
For example, if \co{taken} is not complete, founded semantics gives that
\co{ready\_to\_graduate('mike')} is true and\acmonly{}
\co{ready\_to\_graduate('john')} is undefined, and constraint semantics
gives two models: one with \co{ready\_to\_graduate('john')} being true and
one with it being false.

\mysubsec{Digital circuits---from the most complex to the simplest}

This is Example 25 in~\cite{gelfond2019vicious}, %
finishing the longest span of discussion by building on their Examples 11, %
23, %
and 24 %
that are simpler instances or parts.
It considers a program for propagating binary signals through a digital
circuit that does not have a feedback,
consisting of the following facts
(where \co{input(w,g)} means that wire \co{w} is an input to gate \co{g},
\co{output(w,g)} is similar,
\co{gate(g,'and')} means that gate \co{g} is an \co{and} gate, and
\co{val(w,v)} means that wire \co{w} has value \co{v}):
\begin{code}
  \ieeeonly{}input('w1','g1')\ieeeonly{}  input('w2','g1')\ieeeonly{}  input('w0','g2')
  \ieeeonly{}output('w0','g1')  output('w3','g2')
  \ieeeonly{}gate('g1','and')  gate('g2','and')
  \ieeeonly{}val('w1',0)  val('w2',1)
\end{code}
and a rule:
\begin{code}
  val(w,0) \IF output(w,g) \AND gate(g,'and')
              \ieeeonly{}\AND count \{w: val(w,0), input(w,g)\} \GT 0
\end{code}
Their Example 11 does not have the last fact on each line (i.e., no gate
\co{g2}, input on \co{w0}, output on \co{w3}, and value on
\co{w2}).\footnote{Their Example 11 also reverses the first two hypotheses
  of the rule; this appears to be accidental.}
Their Examples 23 and 24 use simpler instances of the rule to illustrate
their definitions of ``splitting set'' and ``stratification'',
respectively.

First, \co{input}, \co{output}, and \co{gate} as given are certain.  Then,
\co{val} is certain by default, despite that \co{val} is defined using
\co{val} in aggregation, because the dependency is positive---counting with
\GT and with no negation is monotonic.
Therefore, the semantics is simply a least fixed point by using the given
rule, yielding the same result for founded semantics and constraint
semantics: \co{val('w0',0)}, \co{val('w3',0)}, plus the given facts,
consistent with all of Examples 11 and 23--25 in~\cite{gelfond2019vicious}.

\mysubsec{Correlated counts---with different predicate declarations}
\label{sec:correlated-counts}

This is Example 28, the last example, in~\cite{gelfond2019vicious}.  It
considers the following one fact and two rules:
\begin{code}
  p(1)
  p(3) \IF count \{x: p(x)\} \GEQ 2
  p(2) \IF count \{x: p(x)\} \GEQ 2
\end{code}

We have that \co{p} is certain by default, despite that \co{p} is defined
using \co{p} in aggregation, because the dependency is positive---counting
with \GEQ and with no negation is monotonic.
The least fixed point infers \co{p(1)} being true in one iteration, and
ends with \co{p(1)} being true, and \co{p(3)} and \co{p(2)} being false, as
the result of both founded semantics and constraint semantics.
This is the same as the resulting answer set in~\cite{gelfond2019vicious},
but is obtained straightforwardly, not using all possible guesses followed by
sophisticated reducts as in computing answer sets, which, for this example
and for one of the answer sets, considers 9 S-reducts, each containing 3
rules or clauses for a combination of three models each containing 2 of 3
possible facts~\cite{gelfond2019vicious}.

We also consider results under other assumptions, not discussed in~\cite{gelfond2019vicious}.
Suppose that the default is not used, and \co{p} is declared uncertain and
complete.
Then the following completion rule is added, and it does not infer \co{p(3)} or
\co{p(2)} to be false.
\begin{code}
  \ieeeonly{}\NOT p(x) \IF x \NEQ 1 \AND (x \NEQ 3 \OR count \{x: p(x)\} \LT 2)
                   \ieeeonly{}\AND (x \NEQ 2 \OR count \{x: p(x)\} \LT 2)
\end{code}
Founded semantics gives that \co{p(1)} is true, and \co{p(3)} and \co{p(2)}
are undefined.
Constraint semantics gives two models: \co{\{p(1)\}} and
\co{\{p(1),p(2),p(3)\}}.

Suppose that \co{p} is declared uncertain and not complete.  It is
straightforward that \co{p(1)} is true as given, and \co{p(3)} and
\co{p(2)} are left as undefined.  Thus, founded semantics and constraint
semantics are the same as when \co{p} is uncertain and complete.

Supposed that \co{p} is declared uncertain, complete, and closed.  Then the greatest unfounded set
is \co{\{p(3),p(2)\}}, and founded semantics gives that \co{p(1)} is true,
and \co{p(3)} and \co{p(2)} are false.
That is, it makes the last two false, instead of undefined, and is the same
as when \co{p} is certain.  Since there are no undefined values, constraint
semantics has one model: \co{\{p(1)\}}.  This is again the same as
in~\cite{gelfond2019vicious}, but note that 
using \co{p} being certain as above yields this desired result
straightforwardly.
\end{fullversion}

\mysubsec{Company control---a well-known challenge}
\label{sec:company-control}

This is Examples 1.1 and 2.13 in~\cite{faber2011semantics} and is also used in
Example 12 in \cite{gelfond2019vicious}. %
The problem was also discussed repeatedly 
before, e.g.,~\cite{mumick1990,%
kemp1991semantics,%
van92agg,%
ross1997monotonic,%
pelov2007well}, and earlier~\cite{Ceri:Gottlob:Tanca:90}.
It considers a set of facts of the form \co{company(c)}, denoting that \co{c}
is a company, and a set of facts of the form \co{ownsStk(c1,c2,p)}, 
denoting the percentage \co{p} of shares of company \co{c2}
that are owned by company \co{c1}.
It defines that company \co{c1} controls company \co{c2}, denoted \co{controls(c1,c2)}, 
if the sum of the percentages of shares of \co{c2} that are owned 
either directly by \co{c1} or by companies controlled by \co{c1} is more than 50.
\begin{code}
  controlsStk(c1,c1,c2,p) \IF ownsStk(c1,c2,p)
  controlsStk(c1,c2,c3,p) \IF company(c1) 
            \ieeeonly{}\AND controls(c1,c2) \AND ownsStk(c2,c3,p)
  controls(c1,c3) \IF company(c1) \AND company(c3)
            \ieeeonly{}\AND sum \{p,c2: controlsStk(c1,c2,c3,p)\} \GT 50
\end{code}
It introduces \co{controlsStk(c1,c2,c3,p)}, denoting that company \co{c1} controls \co{p} percent of shares of company \co{c3} through company \co{c2}.
It has become a most well-known challenging example for recursion with
aggregation, because it involves aggregation in mutual recursion.

Founded semantics and constraint semantics are\fullonly{ again} straightforward to compute.
First, \co{company} and \co{ownsStk} as given are certain.  Then,
\co{controlsStk} and \co{controls} are certain by default, despite that
\co{controlsStk} and \co{controls} are mutually recursive while involving
aggregation, because \co{controlsStk(c1,c2,c3,p)} holds for only non-negative \co{p}, making the dependency through the comparison positive. Therefore, the semantics
is simply a least fixed point using the given rules, giving the same
result for founded semantics and constraint semantics.  This is the desired result, same as in~\cite{faber2011semantics}.

\mysubsec{Double-win game---for any kind of moves}
\label{sec:double-win}

Consider the following game, which we call the double-win game. %
Given a set of moves, the game uses the following single rule, called
double-win rule, for winning:
\begin{code}
  \ieeeonly{}dwin(x) \IF count \{y: move(x,y) \AND \NOT dwin(y)\} \GEQ 2
\end{code}
It says that \co{x} is a winning position if
the number of positions, \co{y}, such that there is a move from \co{x} to \co{y}
and \co{y} is not a winning position, is at least two.
That is, \co{x} is a winning position if 
there are at least two positions to move to from \co{x} that are
not winning positions.

We created the double-win game by generalizing the well-known win-not-win
game~\cite{LiuSto18Founded-LFCS,LiuSto20Founded-JLC}, 
which has a single rule, stating that
\co{x} is a winning position if there is a move from \co{x} to some
position \co{y} and \co{y} is not a winning position:
\begin{code}
  win(x) \IF move(x,y) \AND \NOT win(y)
\end{code}
One could also rewrite the double-win rule using two explicit positions
\co{y1} and \co{y2} 
and adding \co{y1!=y2}, but this approach does not scale when the count can
be compared with any number, not just 2, and is not necessarily known in
advance.

By default, \co{move} is certain, and \co{dwin} is uncertain but complete.
First, add the completion rule:
\begin{code}
  \ieeeonly{}\NOT dwin(x) \IF count \{y: move(x,y) \AND \NOT dwin(y)\} \LT 2
\end{code}
Then, rename \co{\NOT dwin} to \co{n.dwin}, %
in both the given rule and the completion rule, except the positive occurrence of \co{dwin} in the body of the completion rule, yielding:
\begin{code}
  \ieeeonly{}dwin(x) \IF count \{y: move(x,y) \AND n.dwin(y)\} \GEQ 2
  \ieeeonly{}n.dwin(x) \IF count \{y: move(x,y) \AND \NOT dwin(y)\} \LT 2
\end{code}
Now compute the least fixed point.  Start with %
the base case, in the second rule, for positions \co{x} 
that have moves to fewer than 2 positions;
this infers \co{n.dwin(x)} facts for those positions \co{x}.
Then, the first rule infers \co{dwin(x)} facts for any position \co{x} that
can move to 2 or more positions for which \co{n.dwin} is true.

This process iterates to infer more \co{n.dwin} and more \co{dwin} facts,
until a fixed point is reached, where \co{dwin} gives winning positions,
\co{n.dwin} gives losing positions, and the remaining positions are draw
positions, corresponding to positions for which \co{dwin} is true, false,
and undefined, respectively.

\mysubsec{Over-win game---winning defined over winning}
\label{sec:}

Consider the following game, which we call the over-win game.
Given a set of moves, the game uses the following single rule, called
the over-win rule, for winning:
\begin{code}
  owin(x) \IF count \{y: move(x,y) \AND owin(y)\} \LEQ 2
\end{code}
It says that \co{x} is a winning position if the number of positions,
\co{y}, such that there is a move from \co{x} to \co{y} and \co{y} is a
winning position, is at most two.
That is, \co{x} is a winning position if 
there are at most two positions to move to from \co{x} that are
winning positions.

We created this over-win game from double-win game by removing the negation
on the recursive occurrence of the predicate and reversing the comparison
\GEQ to \LEQ.
That is, winning is now defined over winning, so to speak, instead of over
not winning.  In general, one could of course use any integer in place of
2.
The point is, even though the recursive occurrence of the predicate is no
longer negated, the occurrence of the predicate is still non-positive
because of the reversed comparison.

By default, \co{move} is certain, and \co{owin} is uncertain but complete.
First, add the completion rule:
\begin{code}
  \NOT owin(x) \IF count \{y: move(x,y) \AND owin(y)\} \GT 2
\end{code}
Then, in the given rule, rename the non-positive occurrence of \co{owin}
to \co{\NOT n.owin}, and in the completion rule, rename \co{\NOT owin} to \co{n.owin}, yielding:
\begin{code}
  owin(x) \IF count \{y: move(x,y) \AND \NOT n.owin(y)\} \LEQ 2
  n.owin(x) \IF count \{y: move(x,y) \AND owin(y)\} \GT 2
\end{code}
Now compute the least fixed point.  Start with the base case, in the first
rule, for positions \co{x} that have moves to no more than 2 positions;
this infers \co{n.owin(x)} facts for those positions \co{x}.
Then, the second rule infers \co{n.owin(x)} facts for any position \co{x}
that can move to more than 2 positions for which \co{owin} is true.

This process iterates to infer more \co{owin} and more \co{n.owin} facts,
until a fixed point is reached, where \co{owin} gives winning positions,
\co{n.owin} gives losing positions, and the remaining positions are draw
positions, corresponding to positions for which \co{owin} is true, false,
and undefined, respectively.

\lfcsonly{}

\begin{fullversion}
\section{Experiments}
\label{sec-expe}

We also performed experiments with our new semantics and 
compared with results computed by four well-known systems.  We 
implemented straightforward and incremental least fixed-point computations 
for selected examples in DistAlgo~\cite{Liu+17DistPL-TOPLAS}, an extension of Python.

For comparison with founded semantics, we use XSB~\cite{xsb17}, 
the most well-known system
for computing WFS and that supports negation and aggregation in recursion,
and SWI-Prolog~\cite{wielemaker2011tplp}, which added similar support more recently. 
For comparison with constraint semantics, we use clingo~\cite{alviano2015rewriting} and
DLV~\cite{faber2008design,alviano2017asp},
the most well-known systems for computing SMS and that support negation and
aggregation in recursion.

We show experimental results for the three most complex examples we 
discussed: the company-control problem, the double-win game, and the over-win game.  
For our programs, the default declarations described in Section \ref{sec-examp} are used.
The XSB programs were written by an XSB expert. The SWI-Prolog programs are the same except for using a conjunction with \co{fail} to find all solutions in place of \co{do\_all} that is in XSB but not in SWI-Prolog.
For the company control problem, the DLV and clingo programs are the same, from \cite[Example 1.1]{faber2011semantics}.
For all benchmarks, we compared results from our system with results from the other logic rule engines to check correctness.  We also manually checked correctness of the results for all small benchmarks.  

While the emphasis of our experiments is on semantics and correctness, we also report running time results for the company-control problem.
For the double-win game and over-win game, all other systems either do not compute the desired semantics or were found to compute incorrect results.  XSB and SWI-Prolog have since found and fixed bugs that caused the incorrect results in the double-win game, but developers of these systems also found that
fundamentally different inference would be needed to compute correctly at
all for the over-win game.

Our experiments used five systems: DistAlgo 1.1.0b15, available via \url{https://github.com/DistAlgo/distalgo}, on Python 3.7.0;
XSB 4.0.0, available via \url{http://xsb.sourceforge.net/}; SWI-Prolog 8.5.1, available via \url{https://www.swi-prolog.org/}; 
clingo 5.4.0, available via \url{https://potassco.org/clingo/}; and 
the version of DLV available via \url{https://www.dbai.tuwien.ac.at/proj/dlv/dlvRecAggr/} (accessed 2020-09-21 and 2021-10-01).
That version of DLV supports recursive aggregates, while the current release of DLV ``does not yet contain a full implementation of recursive aggregates'' 
according to the DLV
System website, \url{http://www.dlvsystem.com/dlv/} (accessed 2020-09-21 and 2021-10-01).  We tried to run the company control problem in the current version of DLV (accessed 2021-10-01), but it exited with the error message ``the predicates appearing in the aggregate atom cannot be recursive with the head of the rule''.

\mypar{Company control}
For the company control problem, we compare 
a straightforward least fixed-point computation of our semantics in DistAlgo, and
a more efficient version of that computation obtained using the incrementalization method developed by Liu et al.~\cite{LiuSto09Rules-TOPLAS,Liu+16IncOQ-PPDP,Liu+17DistPL-TOPLAS},
with XSB, SWI-Prolog, clingo, and DLV. 

Our benchmark problem instance generator takes the desired number $N$ of \co{company} facts as input and generates pseudorandom problem instances with $N$ \co{company} facts and $N$ \co{ownsStk} facts.  The generator ensures that the owned percentages of each company sum to at most 100, and that nontrivial transitive control relationships exist.  It accomplishes the latter by pseudorandomly choosing triples \co{c1}, \co{c2}, \co{c3} of companies, giving \co{c1} a controlling share (i.e., more than 50\%) of \co{c2}, and splitting a controlling share of \co{c3} between \co{c1} and \co{c2}.

All measurements were taken on 64-bit Ubuntu 16.04
on a 3.47 GHz Intel Xeon X5690 CPU with 94 GB RAM.
All running times are for computing all inferred facts
as the semantics of the rules and include the time needed to read the \co{company} and \co{ownsStk} facts from a file and print the \co{controls} relation.
Each reported time is an average of CPU times of runs over 10 generated benchmarks of the specified size. %
Our test driver checks and confirms that all five systems produce the same query answers for each benchmark.

\pgfplotstableread[col sep = comma]{
size,DAinc,DAinc-sd,XSBrev-dync,XSBrev-dync-sd,SWIrev-manual,SWIrev-manual-sd,SWIrev-consult,SWIrev-consult-sd,clingo,clingo-sd,XSBrev-dyn,XSBrev-dyn-sd,XSBrev-consult,XSBrev-consult-sd,XSB-consult,XSB-consult-sd,DA,DA-sd,DLV,DLV-sd
100,0.0012,0,0.066,0.0143,0.028,0.0042,0.026,0.0052,0,0,0.061,0.0088,0.096,0.0097,0.134,0.0097,0.8503,0.2098,2.458,0.1169
200,0.002,0,0.061,0.0088,0.031,0.0088,0.034,0.0052,0.003,0.0048,0.075,0.0071,0.114,0.0084,0.27,0.0094,6.6706,2.4264,31.572,0.5767
300,0.0028,0,0.06,0.0067,0.031,0.0057,0.042,0.0079,0.009,0.0032,0.082,0.0063,0.14,0.0094,0.489,0.0152,23.0727,7.6427,219.637,14.9241
400,0.0036,0,0.062,0.0092,0.034,0.0052,0.049,0.0032,0.01,0,0.097,0.0082,0.161,0.0099,0.708,0.1257,50.2752,8.1242,702.539,25.444
500,0.0045,0.0001,0.059,0.0088,0.034,0.0052,0.056,0.0052,0.014,0.0052,0.1,0.0125,0.177,0.0125,0.86,0.0926,111.0051,33.8956,1692.235,57.937
600,0.0052,0,0.07,0.0094,0.038,0.0042,0.061,0.0057,0.019,0.0032,0.114,0.0126,0.211,0.0074,1.103,0.1267,193.3064,38.1546,,
700,0.0061,0.0001,0.068,0.0063,0.037,0.0048,0.069,0.0032,0.02,0,0.12,0.0105,0.225,0.0118,1.351,0.1259,258.1302,61.332,,
800,0.0069,0.0001,0.067,0.0082,0.039,0.0032,0.078,0.0042,0.027,0.0048,0.128,0.0063,0.249,0.0057,1.616,0.1282,374.3979,94.545,,
900,0.0076,0.0001,0.067,0.0048,0.04,0.0047,0.086,0.0052,0.03,0,0.137,0.0095,0.274,0.0117,1.752,0.0308,500.3883,105.2228,,
1000,0.0085,0.0001,0.071,0.0032,0.044,0.0052,0.094,0.0052,0.033,0.0048,0.145,0.0071,0.297,0.0106,2.232,0.1043,633.8809,133.7115,,
1100,,,,,,,,,,,,,,,,,894.40449,309.2197273,,
1200,,,,,,,,,,,,,,,,,1087.7161,291.1303704,,
2000,0.017,0.0002,0.079,0.0099,0.058,0.0042,0.165,0.0071,0.07,0,0.234,0.0126,0.557,0.0116,7.582,0.1478,,,,
3000,0.0251,0.0002,0.09,0.0067,0.07,0.0047,0.243,0.0048,0.109,0.0032,0.319,0.0137,0.808,0.022,16.751,0.3687,,,,
4000,0.0333,0.0001,0.101,0.011,0.087,0.0048,0.317,0.0048,0.146,0.0052,0.405,0.0108,1.006,0.126,29.132,0.3817,,,,
5000,0.0429,0.0001,0.115,0.0053,0.101,0.0032,0.389,0.0088,0.186,0.0052,0.496,0.0084,1.18,0.1912,45.587,0.6764,,,,
6000,0.0509,0.0003,0.122,0.0079,0.116,0.0052,0.435,0.0828,0.218,0.0042,0.575,0.0196,1.266,0.179,64.75,0.7657,,,,
7000,0.0592,0.0003,0.133,0.0095,0.13,0.0067,0.535,0.0085,0.257,0.0048,0.592,0.1139,1.506,0.2198,88.594,1.3365,,,,
8000,0.0674,0.0004,0.138,0.0103,0.14,0.0047,0.576,0.0771,0.291,0.0057,0.634,0.1167,1.612,0.059,119.762,5.1667,,,,
9000,0.0757,0.0005,0.156,0.0165,0.158,0.0092,0.64,0.12,0.337,0.0048,0.705,0.1354,1.727,0.0952,174.873,8.4335,,,,
10000,0.0839,0.0004,0.163,0.0082,0.171,0.0074,0.616,0.1607,0.374,0.0052,0.746,0.0875,1.967,0.1007,218.241,14.4432,,,,
}\ComData

\begin{figure}[t]
  \centering

\newcommand\figCom[1]{
\begin{tikzpicture}%
  \begin{axis}[
    ymin=0, ymax=#1,
    xtick={0,2000,...,10000},
    tick label style={font=\scriptsize},
    restrict y to domain=0:\ifodd#1 2\else2000\fi, %
    xlabel={Number of companies},
    ylabel={CPU time (in seconds)},
    ylabel shift=-1ex,
    mark options={solid},
    legend style={draw=none}, %
    legend style={cells={anchor=west}, at={(0.75,.99)}, anchor=north east},
    legend style={font=\tiny, row sep=-.5ex},
    ymajorgrids=true,
    height=50ex, %
    width=0.51\linewidth,
    ]

\pgfmathparse{( #1 == 1000 ? int(1) : int(0))}
\ifnum\pgfmathresult>0
\addplot[mark=pentagon] table[x=size,y=DLV] {\ComData};
\addplot[mark=square] table[x=size,y=DA] {\ComData};
\else
\fi

\addplot[mark=diamond] table[x=size,y=XSB-consult] {\ComData};
\addplot[mark=o] table[x=size,y=XSBrev-consult] {\ComData};

\addplot[dash pattern=on 5pt off 1.5pt,mark=o] table[x=size,y=XSBrev-dyn] {\ComData};
\addplot[mark=|] table[x=size,y=SWIrev-consult] {\ComData};

\addplot[mark=star] table[x=size,y=clingo] {\ComData};

\addplot[dash pattern=on 3pt off 1pt, mark=|] table[x=size,y=SWIrev-manual] {\ComData};
\addplot[dash pattern=on 3pt off 1pt, mark=o] table[x=size,y=XSBrev-dync] {\ComData};

\addplot[mark=triangle] table[x=size,y=DAinc] {\ComData};

\ifnum\pgfmathresult>0
\legend{DLV, DA, XSB-consult$\!\!\!\!\!\!\!\!\!$, XSBopt-consult$\!\!\!\!\!\!\!\!\!$,
  XSBopt-dyn, SWIopt-consult$\!\!\!\!\!\!\!\!\!$,
  clingo, SWIopt-special$\!\!\!\!\!\!\!\!\!$, XSBopt-dync$\!\!\!\!\!$, DAinc}
\else
\legend{XSB-consult$\!\!\!\!\!\!\!\!\!$, XSBopt-consult$\!\!\!\!\!\!\!\!\!$,
  XSBopt-dyn$\!\!$, SWIopt-consult$\!\!\!\!\!\!\!\!\!$,
  clingo, SWIopt-special$\!\!\!\!\!\!\!\!\!$, XSBopt-dync$\!\!\!\!\!$, DAinc}
\fi

\end{axis}
\end{tikzpicture}\Vex{0}
}
\Hex{-2}\figCom{1000}
\Hex{1}\figCom{1}
  
  \caption{Running time comparison for the company control problem, showing CPU times up to 1000 seconds (left) and zooming in to 1 second to reveal the differences between the fastest systems (right).
  DA and DAinc are the DistAlgo least fixed point computation and the incrementalized version of it, respectively.  XSBopt and SWIopt are XSB and SWI-Prolog, respectively, on the variant with reversed rules.  For XSB and SWI-Prolog, the suffixes after the dash indicate how facts were loaded.
}
  \label{fig-time-comcontrol}
\end{figure}

Figure~\ref{fig-time-comcontrol} shows the running times of the five systems for an increasing number of companies.
For XSB and SWI-Prolog, besides the original rules, we also used an optimized variant with ``reversed'' rules; specifically, the \co{company} hypotheses in the second and third rules are moved to the ends of the rules.  This variant was proposed by an XSB expert after debugging XSB's slow performance on the original rules and led to drastic speedups.
This change to the rules has a similar impact on SWI-Prolog's running time, so we only show its running times on the drastically faster variant.
This change does not affect the running times of the other systems.

Additionally, for loading facts, we experimented with four ways in XSB:
predicate \co{consult/1} for reading a Prolog file, \co{load\_dyn/1} for faster loading of dynamic code, and more specialized versions \co{load\_dync/1} and \co{load\_dynca/1} recommended by an XSB expert.
We also experimented with two ways of loading facts in SWI-Prolog: 
predicate \co{consult/1}, and 22 lines of specialized handwritten code recommended by an SWI-Prolog expert.

Figure~\ref{fig-time-comcontrol}-left shows that our straightforward %
implementation in DistAlgo is much slower than XSB on the original rules but much faster than DLV.  All other XSB, SWI-Prolog, clingo, and DistAlgo programs are too fast to tell apart from 0 seconds with this range of CPU times, regardless of optimizations of rules and methods of loading facts.

Figure~\ref{fig-time-comcontrol}-right shows that our incrementalized implementation in DistAlgo is the fastest; followed by XSB and SWI-Prolog on the reversed rules using \co{load\_dync/1} and using specialized code by hand, respectively, for loading facts; then by clingo; and then by XSB and SWI-Prolog on the reversed rules using other ways to load facts; finally, XSB on the original rules is significantly slower.  Results for XSB using \co{load\_dynca/1} are not shown, because they are basically the same as the results using \co{load\_dync/1}.

For XSB on the original rules, only the results using \co{consult/1} are shown;
the results using \co{load\_dyn/1} and \co{load\_dync/1} are similar to each other and about 15--25\% slower than using \co{consult/1}---this is opposite to those times being faster for the reversed rules because (1) compared to XSB times on the original rules, all ways of loading facts are too fast, being a part of XSB times on the reversed rules, as shown in Figure~\ref{fig-time-comcontrol}-left,
and (2) unlike \co{load\_dyn}, etc., \co{consult/1} statically compiles the facts to support more efficient hashing, showing its advantage for more expensive queries. 

Beyond these performance results, our experiments show that significant expertise is needed to achieve high performance with XSB and SWI-Prolog.

\mypar{Double-win game}
The result for this problem was partly unexpected, because it revealed
that, for many of our benchmarks, both XSB and SWI-Prolog gave incorrect results.
Both clingo and DLV are known to not compute 3-valued semantics.
Therefore, we do not compare with these systems for performance.  
The performance of our DistAlgo implementation for this problem 
is similar to that for the company control problem but is faster
because it involves fewer sets and operations.

For running the double-win game problem in XSB, we were helped by an
XSB expert.  The result is that when querying using \co{dwin(X)} in
XSB to compute true, false, and undefined values for all positions, 
the XSB implementation produced different results than our DistAlgo implementation on many of our benchmarks.  
Some results differ very slightly but some others differ significantly.

However, many logic rule engines also allow the same rules to be directly used
for different queries.  For all benchmarks that XSB produced different
results than ours, when querying using \co{dwin(x)} for each of the
positions \co{x} separately, XSB produced the same true, false, or
undefined values for all positions as our results.

So this experiment helps confirm the correctness of our results and at the
same time revealed a bug in XSB.  Some careful investigation into this bug
indicated that it is a nasty one.  %
After learning that SWI-Prolog added similar support as XSB for computing WFS, 
we also ran tests in SWI-Prolog and found various incorrect results.
Both XSB and SWI-Prolog have since found and fixed bugs that caused these incorrect results.

We also ran the double-win example in both clingo and DLV.  
Both exited with an error message that the rule is unsafe.

Note that 2-valued semantics such as extensions of SMS do not give the desired semantics for this game.  For example, even
on the simple input with only two \co{move} facts, \co{move(1,1)} and \co{move(1,2)}, any such semantics would give \co{\{\}}, i.e., no model, meaning that the game is inconsistent, but the desired result is that \co{dwin(2)} is false, and \co{dwin(1)} is undefined, i.e., \co{2} is a losing position, and \co{1} is a draw position.

\mypar{Over-win game}
The result for this problem was totally unexpected, because 
we found that not only both XSB and SWI-Prolog gave incorrect query results,
but also their developers determined that the inference they do to compute their intended semantics fundamentally cannot compute the correct results.
Both clingo and DLV again cannot compute the desired 3-valued semantics, as for the double-win game.

For XSB and SWI-Prolog, their inference to compute WFS deals with loops through explicit negation, but the recursive appearance of \co{owin} is not explicitly negated.
Additionally, their inference does not correctly handle different comparisons that use the results of the same aggregation.
These issues are being studied by the XSB team, and an entirely
new inference algorithm, based on our founded semantics for
computing WFS, is being developed.  Our DistAlgo programs for
experiments have been used to generate test cases, especially on
larger data, because no other known implementations can
compute correct results.

\end{fullversion}

\mysec{Related work and conclusion}
\label{sec-related}

The study of recursive rules with negation goes back at least to Russell's
paradox, discovered over 120 years ago~\cite{irvine20russell}.
Many logic languages and disagreeing semantics have since been proposed,
with significant complications and challenges described in various survey
and overview articles,
e.g.,~\cite{apt1994negation,RamUll95survey,fitting2002fixpoint,trusz18sem-wbook},
and in works on relating and unifying different semantics,
e.g.,~\cite{dung1992relations,prz94well,schlipf95,lin2004assat,denecker2008logic,hou2010fo,bruynooghe2016first,LiuSto20Founded-JLC}.

Recursive rules with aggregation have been a subject of study soon after
rules with negation were used in programming.  They received an even larger variety
of disagreeing semantics in 20 years, 
e.g.,~\cite{%
kemp1991semantics,%
van92agg,%
sudarshan1993extending,%
consens1993low,%
ross1997monotonic,%
simons2002extending,%
gelfond2002representing,%
marek2004logic,%
marek2004set,%
pelov2007well,%
son2007,%
faber2008design,%
liu2010,%
faber2011semantics,%
ferraris2011%
},
and even more intensive studies in the last few years, e.g.,~\cite{%
gelfond2014vicious,%
shkapsky2015optimizing,%
alviano2015rewriting,%
alviano2015complexity,%
alviano2016evaluating,%
alviano2016non,%
zhang2016characterization,%
gelfond2017vicious,%
zaniolo2017fixpoint,%
alviano2018shared,%
cabalar2018functional,%
gelfond2018vicious,%
cabalar2019gelfond,%
gelfond2019vicious,%
gu19rasql,%
das19bigdata,%
zaniolo19mono,%
wang2020automating,%
vanbesien2021analyzing}, %
especially as they are needed in graph analysis and machine learning
applications.

\fullonly{Aggregation is even more challenging than negation, when used in recursion, because it is more
general.  For example, the count of all values \co{x} for which \co{p(x)}
holds is 0 iff for all \co{x}, \co{p(x)} does not hold, but the count can
be, say, 5, meaning that \co{p(x)} holds for some 5 values,
but does not for the other values, with many possibilities.
Even more different and more sophisticated semantics have been proposed for aggregation than for negation.}

Major related works are as shown in Table~\ref{tab-FSCS}, right column.  
They give disagreeing semantics with each other,
without simple formal explanations for the disagreement, as explained there.
More detailed comparisons with work by Kemp and Stuckey~\cite{kemp1991semantics},
Van Gelder~\cite{van92agg}, Pelov, Denecker, and Bruynooghe~\cite{pelov2007well},
Faber, Pfeifer, and Leone~\cite{faber2011semantics}, Gelfond and 
Zhang~\cite{gelfond2019vicious}, and Hella et al.~\cite{hella1999logics,hella2001logics}
appear in \lfcsonly{}\fullonly{Section~\ref{sec-cmp}}.
Among all, Pelov et al.'s work~\cite{pelov2007well}, recently reworked for Answer Set Programming (ASP)~\cite{vanbesien2022analyzing}, which uses SMS, is notable for proposing a framework that
can be instantiated to extend several prior semantics to handle aggregation.
They develop several separate extended semantics.  
In contrast, our approach
uses simple predicate declarations to capture different assumptions made by different semantics in a unifying single semantics.

Many other different semantics have been studied, all focused on restricted classes or
issues.
The survey by Ramakrishnan and
Ullman~\cite{RamUll95survey} discusses some different semantics,
optimization methods, and uses of recursive rules with aggregation in
earlier projects.
Ross and Sagiv~\cite{ross1997monotonic} studies monotonic aggregation but
not general aggregation.
Beeri et al.\,\cite{beeri1992valid} presents the valid model semantics for
logic programs with negation, set expressions, and grouping, but not aggregation.
Sudarshan et al.\,\cite{sudarshan1993extending} extends the valid model
semantics for aggregation, gives semantics for more programs than Van
Gelder~\cite{van92agg}, and subsumes a class of programs in Ganguly et
al.\,\cite{ganguly1991minimum}, but it is only a 3-valued semantics.
Hella et al.\,\cite{hella1999logics,hella2001logics} study expressiveness
of aggregation operators but without recursion.  Liu et al.\,\cite{liu2010} 
give a %
semantics for logic programs with abstract constraints,
which can represent aggregates, and show that, for positive programs, it agrees with one of Pelov et al.'s semantics~\cite{pelov2007well}.
A number of other works have followed Gelfond and Zhang's line of study for ASP~\cite{cabalar2018functional,cabalar2019gelfond,gelfond2019vicious}.

Zaniolo et
al.\,\cite{ganguly1991minimum,zaniolo1993negation,zaniolo2017fixpoint,gu19rasql,das19bigdata,zaniolo19mono}
study recursive rules with aggregation for database applications,
especially including for big data analysis and machine learning
applications in recent years.  They study optimizations that exploit
monotonicity as well as additional properties of the aggregation operators
in computing the least fixed point, yielding superior performance and
scalability necessary for these large applications.  They discuss insight
from their application experience as well as prior research centering
on fixed-point computation~\cite{zaniolo2017fixpoint}, which essentially
corresponds to the assumption that predicates are certain.

Our founded semantics and constraint semantics for recursive rules with
aggregation unify different previous semantics by allowing different
underlying assumptions to be easily specified explicitly, and furthermore
separately for each predicate if desired.  Our semantics are also fully
declarative, giving both a single 3-valued model from simply least
fixed-point computation and a set of 2-valued models from simply constraint
solving.

The key enabling ideas of simple binary choices for expressing assumptions
and simple lease fixed-point computation and constraint solving are taken
from Liu and
Stoller~\cite{LiuSto18Founded-LFCS,LiuSto20Founded-JLC},
where they present a simple unified semantics for recursive rules with
negation and quantification.\fullonly{
To use the power of founded semantics and constraint semantics in
programming, they propose DA logic~\cite{LiuSto20LogicalConstraints-LFCS,LiuSto21LogicalConstraints-JLC}, for design and analysis
logic, that allows different
assumptions to be specified as one of four meta-constraints, allows the
resulting semantics to be referenced directly, and allows programs to be
easily and modularly specified by using knowledge units.}

Our semantics can be extended for rules with negation in the conclusion, in the same way as in 
\cite{LiuSto18Founded-LFCS,LiuSto20Founded-JLC}.  It can also easily be extended for hypotheses that are equalities or negated equalities between  variables and constants, because such hypotheses are already used in presenting the semantics.
Additionally, DA logic~\cite{LiuSto20LogicalConstraints-LFCS,LiuSto21LogicalConstraints-JLC} can be extended to include aggregation, because aggregations and comparisons are interpreted orthogonally from the additional features in DA logic.

There are many directions for future research, including
additional language features,
efficient implementation methods, and precise complexity guarantees~\cite{LiuSto09Rules-TOPLAS,TekLiu10RuleQuery-PPDP,TekLiu11RuleQueryBeat-SIGMOD} when possible. %

\Vex{2}
\mypar{Acknowledgement}
We would like to thank David S. Warren and Jan Wielemaker for their
excellent help with using XSB and SWI-Prolog.

{\renewcommand{\baselinestretch}{-.1}\footnotesize%
\def\bibdir{../../papers/liubib}        %
\newcommand{\etalchar}[1]{$^{#1}$}

}

\fullonly{\appendix

\mysec{Additional proofs}
\label{sec-proofs}

\begin{myproof}{ of Theorem \ref{thm:consistent}}
  The proof of consistency of the founded model is an extension of the
  corresponding proof by induction for the language without aggregation
  \cite[Theorem 1]{LiuSto20Founded-JLC}; for brevity, we refer to that proof
  as the original proof.
The proof is by induction on the sequence of interpretations constructed in
the semantics by steps that either apply one-step derivability or add negated self-false atoms.  
Two extensions to the
proof are needed to show consistency for the language extended with
aggregation.

To show that steps that apply one-step derivability still preserve
consistency, we extend the original proof to show consistency for comparison literals,
in the sense that $\dc(\pgm,I)$ cannot contain a comparison literal and its 
complement; this result is used, together with the result in the original proof 
that $I$ cannot contain a predicate literal and its complement, to show that
the body of a rule and its negation cannot both be true in $I$.
This result 
follows directly from the definition of
linear-time derivability in Figure~\ref{fig:derivable}: for each pair of
biconditionals for deriving complementary comparisons, the right sides of
those biconditionals are mutually exclusive conditions, i.e., the
conjunction of those two conditions is not satisfiable.

To show that steps that add negated self-false atoms still preserve consistency,
we extend the proof to show that
the extended definition of unfounded set still ensures that none of the
atoms in an unfounded set $U$ for an interpretation $I$ are derivable in
$I \union \neg\cdot U$. This property still holds because the definition
ensures that, for each rule instance $R$ that could be used to derive an
atom in $U$, (1) some hypothesis of $R$ is false in $I$ and hence is false
in $I \union \neg\cdot U$, (2) some positive predicate hypothesis of $R$ is
in $U$ and hence is false in $I \union \neg\cdot U$, or (3) some comparison
hypothesis of $R$ is false in $I \union \neg\cdot U$.  Note that these
three cases correspond to the three cases in the extended definition of
unfounded set.

For consistency of constraint semantics, note that constraint models are
required to be interpretations, which are consistent by definition.
\end{myproof}

\begin{myproof}{ of Theorem \ref{thm:model}}
The proof that $\founded(\pgm)$ is a model of $\pgm$ and $\cmpl(\pgm)$ is
an extension of the corresponding proof for the language without
aggregation \cite[Theorem 2]{LiuSto20Founded-JLC}; for brevity, we refer to that proof
as the original proof.  The original proof relies on the result that the body of a ground instance $A~\IF~B$ of a rule in $\cmpl(\pgm)$ cannot become true in $\addneg$ for an SCC $C$, i.e., as a result of adding negative literals for certain predicates in $C$ to the interpretation, where $C$ is the SCC containing the predicate $Q$ in the conclusion $A$.  This result is shown in the original proof by proving by contradiction that $B$ cannot contain a negative literal for a certain predicate in $C$.  To show that this result holds in the language with aggregation, we prove by contradiction that adding those negative literals to the interpretation cannot cause a comparison atom in $B$ to become derivable.

We suppose that $B$ contains a comparison atom $A$ that becomes derivable and therefore true in $\addneg$ for $C$ and show a contradiction.  Since $A$ becomes true in $\addneg$ for $C$, $A$ must contain a non-positive occurrence of a certain predicate $P$ in $C$.  $P$ and $Q$ are in the same SCC $C$, so $P$ must be defined, directly or indirectly, in terms of $Q$.  Since $P$ is certain and is defined in terms of $Q$, $Q$ must be certain.  Since $Q$ and $P$ are defined in the same SCC $C$, and $Q$ depends non-positively on $P$, $Q$ has a circular non-positive dependency, so $Q$ must be uncertain, a contradiction.

Constraint models are 2-valued models of $\cmpl(\pgm)$ by definition.
Any model of $\cmpl(\pgm)$ is also a model of $\pgm$, because $\pgm$ 
is logically equivalent to the subset of $\cmpl(\pgm)$ obtained by 
removing the completion rules added by $\addinv$.
\end{myproof}

\newcommand{\UA}{{\it UA}}

\begin{myproof}{ of Theorem \ref{thm:uncertain-certain}}
First, we show that $\founded(\pgm)$ is 2-valued for predicates in $C$.  Let $RS$ be the set of all ground instances of combined rules and completion rules in $\cmpl(\pgm)$ for predicates in $C$.  Note that every positive literal and negative literal for every predicate in $C$ appears as the conclusion of at least one rule in $G$ (this holds even if rules in $\pgm$ contain conclusions of forms such as \co{p(x,x)} or \co{p(x,0)}, due to the fresh variables and existential quantifiers introduced by $\comb$).  Let $\UA$ be the set of ground predicate atoms for predicates in $C$ that are undefined in $\founded_0(\pgm)$.  For each atom $A$ in $\UA$, since the predicate in $A$ is complete and $A$ is not true or false in $\founded_0(\pgm)$, (1) for every rule $R$ in $RS$ with conclusion $A$, some hypothesis of $R$ is false or undefined in $\founded_0(\pgm)$, and (2) for some rule $R$ in $RS$ with conclusion $A$, some hypothesis of $R$ is undefined in $\founded_0(\pgm)$.  Since all predicates in SCCs that precede $C$ are certain, these undefined hypotheses must be atoms in $\UA$ or their negations, or comparisons whose outcomes depend on the truth value of atoms in $\UA$ or their negations.  

Define a dependence relation $\rightarrow$ on $\UA$ by: $B \rightarrow A$ if some rule $R$ in $RS$ with conclusion $A$ has an undefined hypothesis that is $B$, $\neg B$, or a comparison whose outcome depends on the truth value of $B$.  The previous observation implies that, for every $A$ in $\UA$, there exists $B$ in $\UA$ such that $B \rightarrow A$.  Since $\UA$ is finite, this implies that every atom in $\UA$ is in a $\rightarrow$-cycle.  Since predicates in $C$ do not have circular non-positive dependency, this implies that all dependencies in the cycles are positive.
These observations, together with all predicates in $C$ being closed, imply that the predicate literals in every cycle, and hence all atoms in $\UA$, are in $\selffalse_\pgm(\founded_0(\pgm))$.  Note any comparisons that give rise to the dependencies in the cycles becomes false when all atoms in $\UA$ are false (and hence satisfy clause (3) in the definition of unfounded set), because the comparison must be true or false when all literals it depends on are true or false, and setting literals on which it has positive dependence to false cannot make the comparison become true.  This implies that $\founded(\pgm)$ contains the negations of all literals in $\UA$.  Therefore, $\founded(\pgm)$ is 2-valued.

Next, we show that $\founded(\pgm)=\founded(\pgm')$.  For each predicate $P$ in $C$, the two programs contain equivalent rules for adding positive literals for $P$ to the founded model, because the combined rule for $P$ in $\pgm$ is logically equivalent to the original rules for $P$ in $\pgm$, so $\founded(\pgm)$ and $\founded(\pgm')$ contain the same positive literals for $P$.  Since both models are 2-valued for $P$, they also contain the same negative literals for $P$.

Finally, we show that $\constraint(\pgm)=\constraint(\pgm')$.  We consider the conditions in the definition of $\constraint$, in turn.

Consider the first condition, namely, $\founded(\pgm)\subseteq M$.  It is equivalent for the two programs, because they have the same founded model.

Consider the second condition, namely, $M$ is a model of $\cmpl(\pgm)$.  It differs for the two programs in that $\cmpl(\pgm)$ contains combined rules and completion rules for predicates in $C$, while $\cmpl(\pgm')$ contains the original rules in $\pgm$ for predicates in $C$.  Since $\founded(\pgm)=\founded(\pgm')$, Theorem \ref{thm:model} implies $\founded(\pgm)$ is a model of both $\cmpl(\pgm)$ and $\cmpl(\pgm')$.  Since $\founded(\pgm)$ is 2-valued for predicates in $C$ and all predicates on which they depend, it is 2-valued for all predicates used in those rules.  Therefore, every interpretation $M$ that is a superset of $\founded(\pgm)$ contains the same predicate literals as $\founded(\pgm)$ for all predicates used in those rules and hence is also a model of both $\cmpl(\pgm)$ and $\cmpl(\pgm')$.  Since $M$ is (by the first condition) a superset of $\founded(\pgm)$, and it is a model of both $\cmpl(\pgm)$ and $\cmpl(\pgm')$.  Thus, the second condition is equivalent for the two programs.

Consider the third condition, namely, there does not exist a non-empty subset $S$ of $M\setminus\founded(\pgm)$ containing only positive literals for closed predicates and such that $S = \selffalse_\pgm(M\setminus S)$.  Since $\founded(\pgm)$ is 2-valued for predicates in $C$, and $S \subseteq M\setminus\founded(\pgm)$, $S$ cannot contain literals for predicates in $C$; furthermore, this and the disjointness condition in the definition of unfounded set imply that $\selffalse_\pgm(M\setminus S)$ cannot contain literals for predicates in $C$.  The same conclusions hold for $\pgm'$, since $\founded(\pgm)=\founded(\pgm')$.  Since $S$ and $\selffalse_\pgm(M\setminus S)$ do not contain literals for predicates in $C$, and $\pgm$ and $\pgm'$ have the same rules and declarations for predicates not in $C$, $\selffalse_\pgm(M\setminus S) = \selffalse_{\pgm'}(M\setminus S)$.  Thus, the third condition is equivalent for the two programs.
\end{myproof}

}

\end{document}